\documentclass[twocolumn]{aastex6}
\usepackage{graphicx}
\usepackage{amsmath}
\usepackage{amsthm}
\usepackage{amssymb}
\usepackage{enumerate}
\usepackage{natbib}
\usepackage{array}
\usepackage{hyperref}
\usepackage{rotating}
\usepackage{multirow}
\usepackage{color}
\hypersetup{
colorlinks= true,
allcolors = blue
}

\shorttitle{Short Timescale Variability in the Ultraviolet}
\shortauthors{de la Vega \& Bianchi}

\begin{document}

\def\fchar{$f_{\rm{char}}$} 
\def\gphoton{{\tt gPhoton}} 

\title{Searching for Short-Timescale Variability in the Ultraviolet with the GALEX {\tt gPhoton} Archive I.: Artifacts and Spurious Periodicities}
\author{Alexander de la Vega}

\author{Luciana Bianchi}
\affil{Department of Physics and Astronomy, Johns Hopkins University, Baltimore, MD 21218, USA}


\begin{abstract}

In order to develop and test a methodology to search for UV variability over the entire GALEX database
down to the shortest time scales,
we analyzed time-domain photometry of $\sim5000$ light curves of $\sim300$ bright $(m_{\rm{FUV}},  m_{\rm{NUV}} \leq 14)$ and blue $(m_{\rm{FUV}} - m_{\rm{NUV}} < 0)$ GALEX sources. 
Using the \gphoton \ database tool, we discovered and characterized instrumentally-induced
variabilities in time-resolved GALEX photometry, which may severely impact automated searches for short-period variations. 
The most notable artifact is a quasi-sinusoidal variation mimicking
light curves typical of pulsators, seen occasionally in either one or both detectors, with amplitudes of up to 0.3 mag and periods  corresponding to the periodicity of the spiral dithering pattern used during the observation (P$\sim$120 sec). Therefore, the artifact may arise from small-scale response variations. 
Other artifacts include visit-long ``sagging'' or ``hump'' in flux,  occurring when the dithering pattern is not a spiral,  or a one-time change in flux level  during the exposure.  
These instrumentally-caused variations were  
not reported before, and are not due to known (and flagged) artifacts such as hot spots, which can be easily
eliminated.
To characterize the frequency and causality of such artifacts, we  apply Fourier transform analysis to both light curves and 
dithering patterns, and examine 
whether artificial brightness variations correlate with visit  or instrumental parameters. 
Artifacts do not correlate with source position on the detector. 
We suggest methods to identify artifact variations and to correct them when possible. 

\end{abstract}

\keywords{stars: variables -- stars: oscillations (including pulsations) -- (stars:) novae, cataclysmic variables -- (stars:) white dwarfs -- ultraviolet: stars -- methods:analysis -- astronomical databases:  surveys}


\section{Introduction}


The Galaxy Evolution Explorer \citep[{\it GALEX},][]{martin05}, a NASA Small Explorer orbiting observatory, surveyed the sky in the ultraviolet (UV) from 2003 to 2013. 
Two micro-channel plate (MCP) photon-counting detectors, one in the far-UV (FUV, range 1350 -- 1750 \AA, $\lambda_{\textrm{eff}} = 1516$ \AA ) 
and one in the near-UV (NUV, range 1750 -- 2750 \AA, $\lambda_{\textrm{eff}} = 2267$ \AA), 
each with a 1.25 degree field-of-view (FOV), recorded cascades of electrical signals (known as `events') from photons landing on the MCPs with a time resolution of 5 milliseconds. 
Photon positions and arrival times were recorded and integrated by the mission pipeline over exposure times at each observation or ``visit,''
typically ranging from 150 seconds to 1500 - 1800 seconds (\citealt[(hereafter M07)]{morrissey07}; \citealt{bianchi09, bianchi11a}).
A $\sim 1$ arcmin spiral dither pattern with a cycle nearly two minutes long was used in exposures in the
Medium Imaging and Deep Imaging Surveys (MIS and DIS, respectively; exposures were
typically longer than 1000 sec), but almost never for the All-sky Imaging Survey (AIS; $\sim150$ sec exposures). 
This dithering was adopted to maximize photometric quality by averaging over pixels with different response 
and to avoid detector ``fatigue'' from prolonged exposure of some areas to high count rate events. 

Most earlier studies of variability in the UV with GALEX have used the pipeline-provided photometry integrated over separate observations 
\citep{welsh06, welsh07, welsh11, wheatley12, gezari13,conti14}.   
The full potential of the high temporal resolution achievable through {\it GALEX} photometry, however,  has been hardly explored to date
\citep{robinson05, welsh06, welsh07, welsh11, wheatley08, wheatley12, browne09}, 
because the full time-resolved photon lists have been not publicly accessible until recently (see e.g., \citet{bianchi14l} for a recent review summary of the GALEX mission).   

Recently, \citet[][hereafter M16]{million16} released the first database 
tool enabling time-resolved GALEX photometry, \gphoton.
This tool, however, has only been used to examine single objects \citep[e.g.][]{davenport18} 
or specific stellar populations in varying sample sizes \citep[e.g.][]{boudreaux17, tucker18} and on time scales of 15 - 30 sec. 
The capabilities of \gphoton \ on shorter timescales are not thoroughly known.

This paper presents the first comprehensive analysis of the short-term variability detection capabilities using \gphoton. The analysis revealed a number 
of instrumentally-induced variations in the source count rate, which were not previously reported and must be taken into account in any study using \gphoton.

This paper is organized as follows. In Section \ref{sec:source} we define our sample of sources.
Sections \ref{sec:photom} and \ref{sec:analysis} outline our methods to perform time-resolved photometry and search for variability within light curves
in our sample, respectively. In Section \ref{sec:artifacts} we describe instrumentally induced variability and examine 
whether there are correlations with observational parameters. We develop and test a methodology to detect and remove artifacts in Section \ref{sec:detect_char}.
In Section \ref{sec:conclusion} we summarize and conclude.
We use AB magnitudes throughout this paper. 
All light curves shown in this study do not include aperture corrections.

\begin{figure*}
\centering
\includegraphics[width=5in]{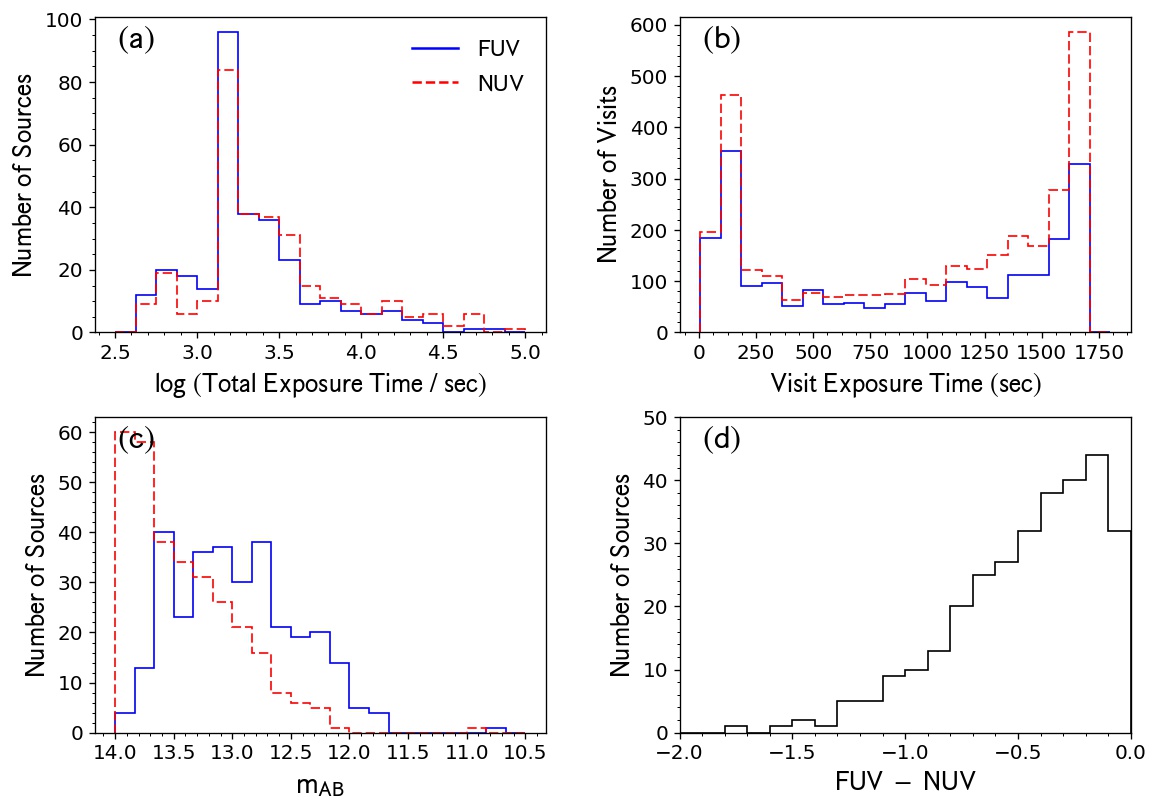}
\caption{Our analysis sample of 304 sources, comprising 5021 visits.
Distribution of 
(a) total exposure time in FUV and NUV,
(b) visit exposure time in each band,
(c) apparent magnitude in each band, and
(d) FUV-NUV color.}
\label{fig:sample}
\end{figure*}

\section{The Source Sample}
\label{sec:source}

In order to develop and test the methodology to search for UV variability over the entire {\tt gPhoton} database down to the shortest
time scales, we selected an initial sample of sources with high count rate in both FUV and NUV. 
Our sample of bright stars is extracted from the General Release 6 and  7 (GR6+7) GALEX merged catalog (MCAT). 
We select sources with $m_{\rm{FUV}} - m_{\rm{NUV}} < 0$ that are brighter than 14th mag in both FUV and NUV, 
and have magnitude error $< 0.1$ mag in each band. To increase the probability of detecting possible periodic variations,
we restrict our sample to sources that have a total exposure time of at least 500 seconds. 
This search yields 350 sources, observed in a total of 
4556 visits, 3186 visits exposed in both FUV and NUV and 1370 visits in which only the NUV detector was on. 
The GALEX MCAT includes multiple entries for the same source.
We identified 31 duplicate sources (totaling 861 exposures in FUV and NUV and 409 exposures in just NUV) and thus have a sample of 319 unique sources.
We then eliminated 15 sources in extended objects, totaling 118 exposures in the FUV and NUV and 21 in NUV only,
and analyzed a final sample of 304 sources, observed in 2207 visits in both bands and 940 in only NUV.
In Figure \ref{fig:sample} we show the distribution of total exposure time, 
magnitude, FUV-NUV color, and per-visit exposure time for both bands. 
Visit exposure times range from 90 seconds to 1750 seconds.

We performed time-resolved photometry by integrating source counts on short time bins during each observation, 
using the \gphoton \ software suite (M16) in the whole database. We have chosen sources with high count rates
so we could test our methodology with very short time integrations, therefore many
magnitudes are close to or brighter than the 10\% non-linearity cutoff (13.73/13.85 AB mag in FUV/NUV) 
measured by \citet{morrissey07}. 
Most sources are well below the GALEX count rate safety limits (9.57/8.89 AB mag in FUV/NUV), as shown in Figure \ref{fig:sample}, panel (c),
and only a handful have FUV magnitudes brighter than where the non-linearity becomes severe (around 12th magnitude in FUV) or unrecoverable (see Figure 8 in \citet{morrissey07}).
Thus, this work explores both the non-linear regime and the bright (still linear) range. 
Seven sources have average count-rate in the linear regime (below the 10\% non-linearity rolloff) observed in 22 FUV visits and 38 NUV visits.  
More sources have average count-rates in the linear regime only in some visits: 272 FUV visits of 107 sources and 
825 NUV visits of 185 sources, but are above the non-linearity rolloff in other visits. 

For the analysis, we will only consider measurements of targets when the 
average position of the target during the visit  is within the central 1$^{\circ}$ of the 
GALEX field, to avoid rim artifacts and distortions affecting sources close to the edge \citep{bianchi11a}. 
The initial limit of {\tt fov-radius} $\leq 33$',  the default value for \gphoton, is used to vet visits that suffer from rim artifacts. 
This eliminates 3 exposures in which the target is farther than 33' from the field center during the entire exposure, 
while for 85 exposures the target is occasionally exceeding this radius 
due to the dithering pattern during the observation. \gphoton \ does not return measurements for these bins. 

We use a time resolution of 5 seconds as a starting point.
Typical background count rates of $3\times 10^{-4} \ \textrm{counts \ s}^{-1} \ \textrm{pixel}^{-1} \ \left(10^{-3} \ \textrm{counts \ s}^{-1} \ \textrm{pixel}^{-1}\right) $
\footnote{See \\ \href{https://asd.gsfc.nasa.gov/archive/galex/FAQ/counts\_background.html}{https://asd.gsfc.nasa.gov/archive/galex/FAQ/counts\_background.html} }
amount to roughly 1\% (5\%) in the FUV (NUV) source counts at the sample limiting magnitude (14 mag).
We further limit our analysis to time bins with an effective exposure time of at least 75\% the time bin size 
to avoid spurious variations caused by underexposed events.
In 149 exposures, effective integration times are shorter than 75\% of the time bin in all measurements, due to 
the high count rates and frequent dead-time corrections. 
These cuts leave 5021 light curves in our sample, from 2141 visits in FUV and NUV and 838 in NUV only.

In Table \ref{tab:sources} we list the sources of the culled sample. We give the MCAT GALEX identifier, right ascension, declination, MCAT magnitudes, total exposure time, number of visits,
identifier as resolved by the SIMBAD database \citep{wenger00}, and the total change in magnitude for the source, $\Delta \textrm{mag}$ defined in Section \ref{sec:analysis}.

\begin{sidewaystable*}[h!]
    \centering
    \caption{The sources of the sample.}
    \begin{tabular}{cccccccccccc}
    \hline
 GALEX {\tt objid} & R. A. & Decl. & FUV mag & NUV mag & \multicolumn{2}{c}{Total exposure time (sec)}  &  \multicolumn{2}{c}{Number of visits} & SIMBAD Identifier & \multicolumn{2}{c}{$\Delta \textrm{mag}$} \\
  & (deg.) & (deg.) & (AB) & (AB) & FUV & NUV & FUV & NUV & & FUV & NUV \\
    \hline
3045596768909664637 & 56.220652 & -52.952540 & $12.974 \pm 0.003$ & $13.350 \pm 0.002$ & 1021.050 & 1021.050 & 3 & 5 & HD23722 & 0.237 & 0.424 \\
2631476709424110117 & 18.788788 & 30.842748 & $13.295 \pm 0.002$ & $13.301 \pm 0.001$ & 1094.000 & 4342.600 & 7 & 7 & TYC2291-741-1 & 7.426 & 0.353 \\
2420511214379470466 & 127.414863 & 47.373656 & $13.670 \pm 0.002$ & $13.729 \pm 0.001$ & 1184.000 & 2885.000 & 2 & 3 & \ldots & 0.392 & 0.212 \\
2938354802782898759 & 220.322268 & 1.624510 & $13.026 \pm 0.001$ & $13.812 \pm 0.001$ & 1187.550 & 2480.250 & 4 & 6 & PG1438+018 & 0.556 & 0.797 \\
2556709918869569243 & 15.483353 & -74.590233 & $13.574 \pm 0.003$ & $13.797 \pm 0.001$ & 1189.150 & 3838.150 & 4 & 9 & \ldots & 0.499 & 0.495 \\
3377913163288680540 & 276.070745 & -49.853054 & $12.513 \pm 0.001$ & $12.919 \pm 0.001$ & 1201.800 & 1201.800 & 9 & 9 & HD168804 & 2.044 & 1.881 \\
3207655986952143500 & 134.609080 & -5.813631 & $12.560 \pm 0.001$ & $13.125 \pm 0.001$ & 1209.850 & 1742.150 & 6 & 10 & HD76816 & 1.288 & 0.920 \\
3120961693923939396 & 12.766740 & -19.999564 & $13.576 \pm 0.003$ & $13.823 \pm 0.002$ & 1220.550 & 1220.550 & 3 & 3 & GD656 & 0.436 & 0.196 \\
2923436629017236386 & 45.141489 & -11.435283 & $12.853 \pm 0.002$ & $12.989 \pm 0.001$ & 1265.600 & 1265.600 & 3 & 3 & Feige30 & 0.760 & 0.547 \\
     \hline
    \end{tabular}
    \label{tab:sources}
\end{sidewaystable*}

\section{Photometry}
\label{sec:photom}

We use \gphoton \ Version 1.27.2. We calculate time-resolved photometry with the {\tt gAperture} tool using aperture radii of 15 and 25'', to account for the FWHM of 4.2'' and 5.3''  
of the GALEX point-spread function in FUV and NUV, respectively, and examine the effects of aperture corrections, non-linearity and saturation.

In the analysis of light curves that follows, we will only use data points 
(i) with effective exposure $\geq 75$\% of the time bin, 
(ii) with distance from the field center $\leq30$', (iii) not affected by hotspots or pixels with response $< 0.7$ in the source or 
background integration area. We will refer to this restricted data set as the ``clean sample".
We illustrate the reason for these cuts below. 

\subsection{Background estimation}
\label{subsec:bckg}

Background estimation for each photometric measurement is computed by {\tt gAperture} by integrating
the flux within a user-specified annulus surrounding the source and scaling it to the aperture area (M16).
For the 15'' aperture photometry we utilize inner and outer annulus radii of 30'' and 45'' for background subtraction.
For the 25'' aperture photometry we use inner and outer annulus radii of 35'' and 50''.  
For comparison, we also used a previous version of \gphoton \ (Version 1.26.2) which provided 
a `swiss-cheese' background method in which photon events from nearby stars (identified in the MCAT catalogue) were masked and 
excluded from background calculations. This option is not available in later versions of \gphoton \ (M16).
Another background estimation method in \gphoton \ involves scaling the local sky background from the pipeline-produced background
image of each field for each visit to the aperture area. The current annulus background technique in \gphoton \ can account for variable sky background 
in each visit, unlike the method using the MCAT catalogue, but suffers from the possible presence of sources in the annulus, which
must be excluded from the background annulus.

\begin{figure*}
\centering

\begin{tabular}{cc}
\includegraphics[width=2.7in]{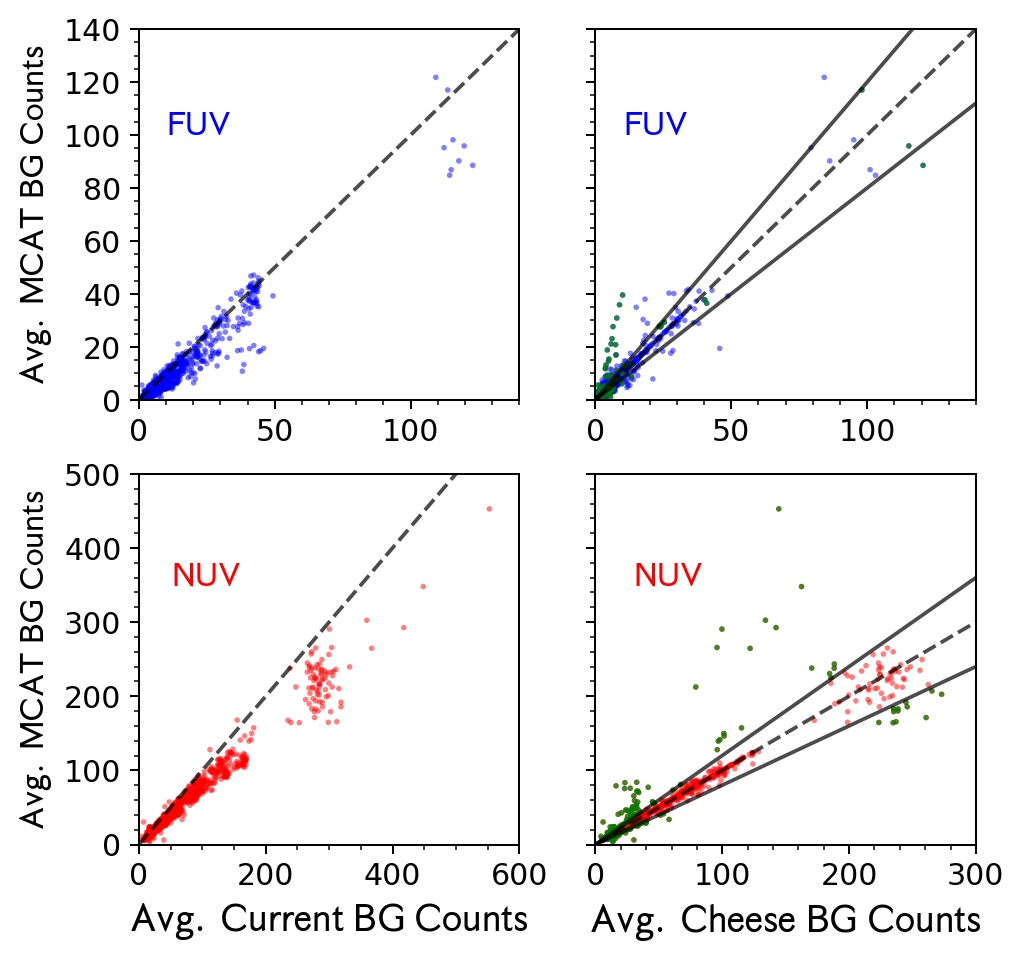}
&
\includegraphics[height=2.25in]{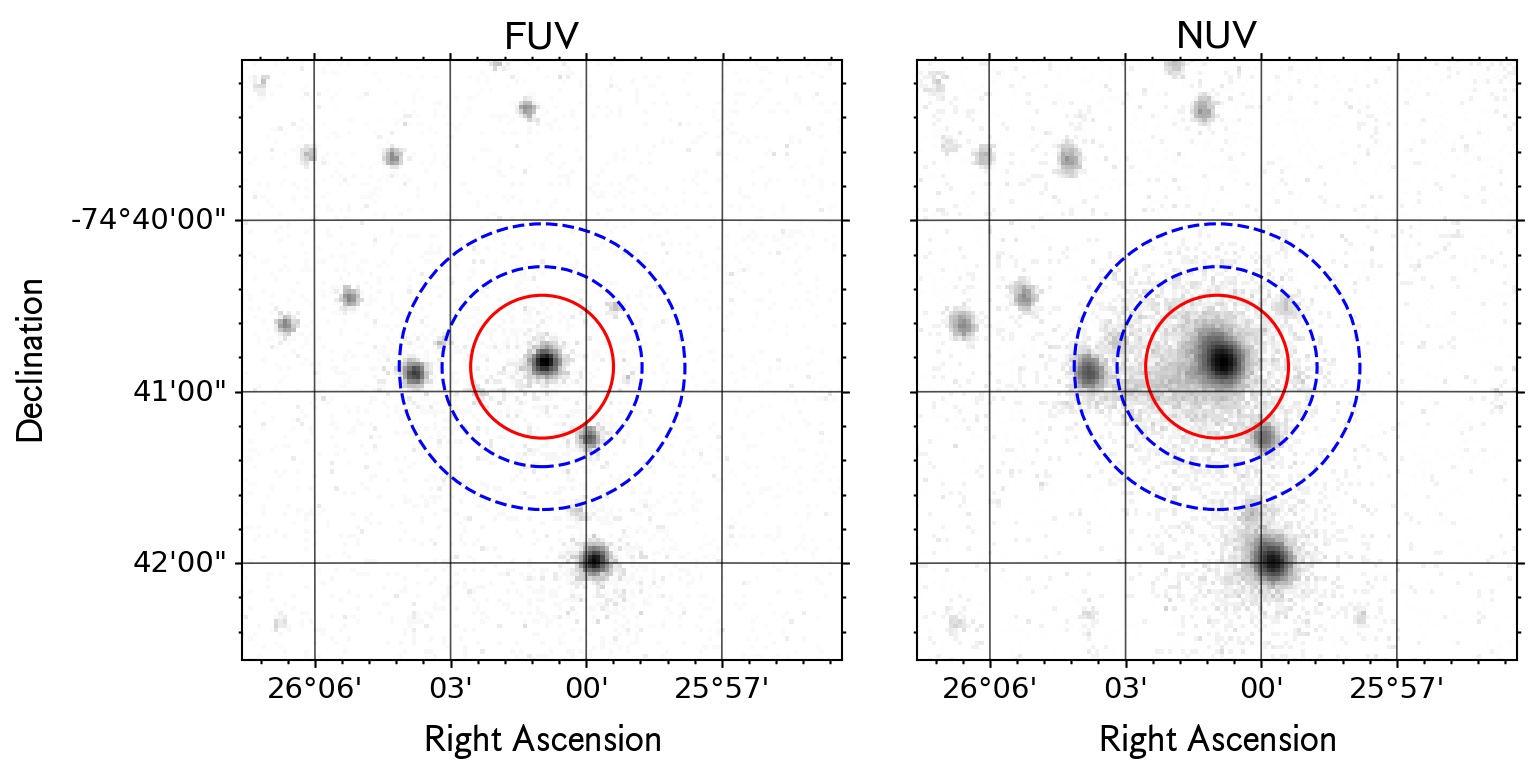}
\end{tabular}

\caption{ Left top panels: Scatter plots of average current background estimation in \gphoton \ versus MCAT background estimation (left) and versus `swiss-cheese' 
background estimation (right) for FUV visits.  
Left Bottom panels: Same as top panel, but for NUV visits. Dashed lines indicate a line with zero intercept and slope 1. Solid lines with zero intercept
and slopes of 1.2 and 0.8 are plotted to identify outliers from the one-to-one correspondence. 
Exposures with MCAT background 20\% greater than or lesser than the swiss cheese background in NUV are colored green in both scatter plots on the right.
The right panels show a source ({\tt objid} 3069733248162596715) with faint, neighboring sources within the background annulus. 
North is up and East is left, and the source is shown in a log stretch. 
Red circles indicate the 25'' aperture we use and blue dashed circles indicate our background annulus radii of 35'' and 50''.}
\label{fig:bckg_compare}
\end{figure*}

In Figure \ref{fig:bckg_compare} we compare results from different background estimation methods mentioned above. We compared 2028 visits in FUV and 
2638 visits in NUV from Version 1.26 of \gphoton \ and compute the average background counts per visit 
according to the `swiss-cheese' method, the MCAT method (`Old BG Counts' in the figure) and the currently implemented (\gphoton \ Version 1.27.2)
annulus background method. The presently-implemented background estimation from a local annulus 
is roughly consistent with the MCAT -- based background estimate, but when compared to the `swiss-cheese' background, it exhibits more scatter, with
discrepancies occasionally larger than 20\%. The difference highlights the need to remove sources that 
fall within the annulus from the background estimate, or to adjust the annulus to exclude nearby sources, if possible, to avoid overestimating the background.
This is the case for objects such as source ID = 3069733248162596715, which has nearby sources 
(the right-hand panels in Figure \ref{fig:bckg_compare}). 

We further restrict our analysis to visits with MCAT-based background estimates within 20\% of those estimated 
using the `swiss-cheese' method. For these visits we use the currently-implemented background estimates in which the flux within 
the annulus is scaled to the aperture area. We remove visits with background
estimates differing by more than 20\% from the `swiss-cheese' method estimates. 


\subsection{Photometric error estimates}

We compute photometric errors as follows. We extract from the output \gphoton \ photometry (i.e. output from the {\tt gAperture} routine)
the total counts per time element as well as the background 
counts. The background counts $N_{\textrm{bckg}}$ are scaled to the area of the aperture and the error in counts per time element:

\begin{equation}
\sigma_{\textrm{source (counts)}} = \sqrt{ N_{\textrm{aper}} + N_{\textrm{bckg}} \times \frac{r_{\textrm{aper}}^2}{r_{\textrm{out}}^2 - r_{\textrm{in}}^2} },
\end{equation}

\noindent where $ N_{\textrm{aper}}$ is the total counts per time element within the aperture, 
$N_{\textrm{bckg}}$ the total background counts in the annulus per time element, $r_{\textrm{aper}}$
the aperture radius, and $r_{\textrm{in}}$ and $r_{\textrm{out}}$ the inner and outer annulus radii, respectively. 
The magnitude error is computed as:

\begin{equation}
\sigma_{\textrm{source (mag)}} = 2.5 \log \left( 1 + \frac{\sigma_{\textrm{source (counts)}}}{ N_{\textrm{aper}} - \left( N_{\textrm{bckg}} 
\times \frac{r_{\textrm{aper}}^2}{r_{\textrm{out}}^2 - r_{\textrm{in}}^2}\right) } \right).
\end{equation}

\subsection{Stability and aperture correction}

\begin{figure*}
\centering

\begin{tabular}{cc}
\includegraphics[height=2in]{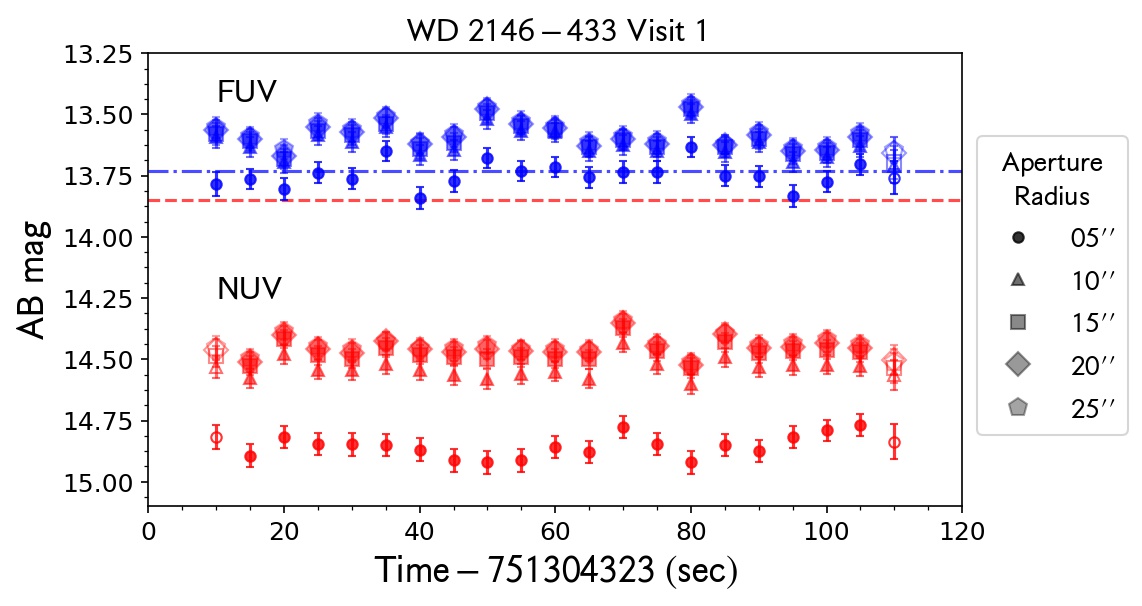}
&
\includegraphics[width=2.7in]{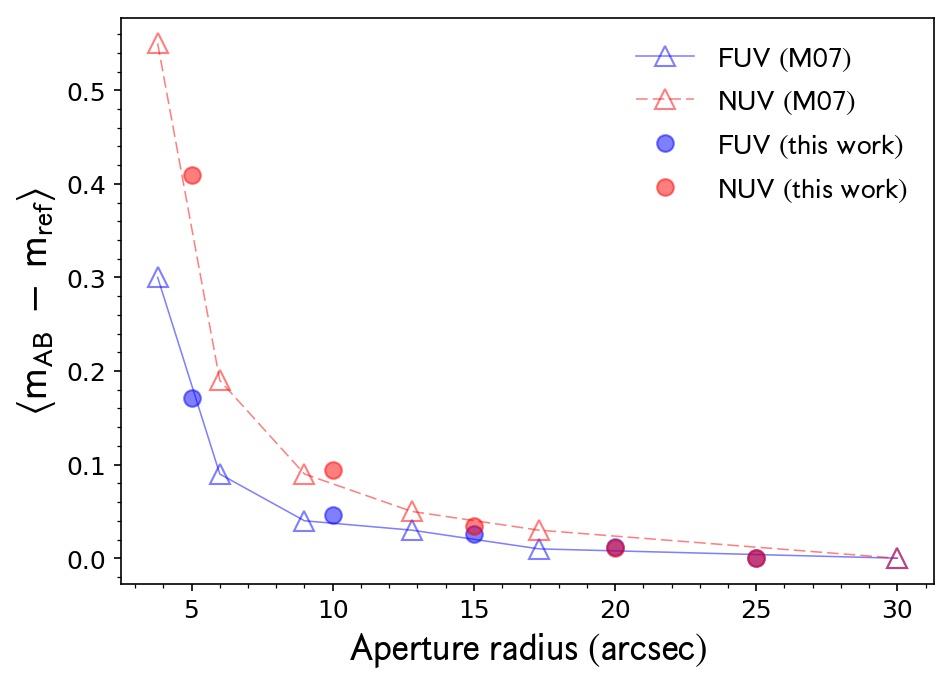}
\end{tabular}

\caption{Light curve of WD 2146-433 in FUV and NUV for five apertures (left panel). 
Blue dash-dot and red dashed lines indicate the 10\% non-linearity rolloff in FUV and NUV in M07, respectively.
Average ``curve of growth'' in the FUV and NUV (right panel). 
The curve of growth in Fig. 5 in M07 is shown for comparison (triangles).
Significant variations in the photometry with apertures with radius $< 15''$ are recovered in all larger apertures. 
}
\label{fig:aper_corr}
\end{figure*}

While the astrometry in both GALEX and the \gphoton \ coordinate reconstruction is reported to be more accurate than 1'' ($\alpha, \delta < 1.0''$), 
spurious variability could result if the dithering pattern was not compensated for with high precision during the exposure. 
As a test of stability against drifts, we show
in the left panel of Figure \ref{fig:aper_corr} the light curve for the white dwarf WD 2146-433 for five different aperture radii, from 5'' to 25'', 
in both FUV and NUV. Fluctuations in both bands using the 5'' aperture radius are recovered consistently in all larger apertures.
The right panel of Figure \ref{fig:aper_corr} shows the average ``curve of growth'' with different apertures for one visit of WD 2146-433.
The magnitude for the 25'' aperture is used as a reference for the curve of growth. The curve of growth for WD 2146-433
agrees well with Fig. 4 in M07, which used the white dwarf LDS749b as a reference standard.

\begin{figure}
\centering

\includegraphics[width=2.75in]{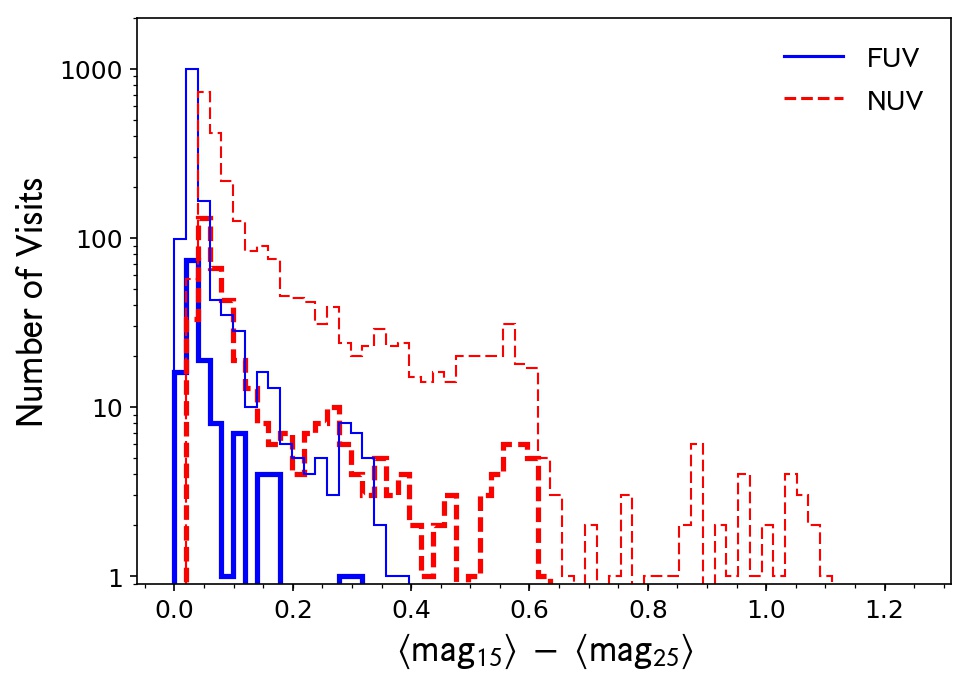}

\caption{
Histogram of average difference in magnitude between 15'' and 25'' radius apertures with  
average count rate in the linear regime (thick lines) and all visits (thin lines). 
We attribute the long tail in the NUV aperture differences to saturated sources.
}
\label{fig:aper_corr_tot}
\end{figure}

We quantify the aperture correction between the 15'' and 25'' radius apertures in our sample by computing the average magnitude
difference in each visit in the clean sample.
In Figure \ref{fig:aper_corr_tot} we show the distribution of this average magnitude difference for all visits.
The majority of visits have an average difference between the 15'' and 25'' radius apertures $<0.1$ mag, in good agreement with the curve of growth 
shown in M07.
For some sources the difference between the apertures approaches 1 mag; these sources are often bright (mag $<12$) 
have unusually wide (radius $\sim 45''$) ``skirts'' in NUV and much diffuse light in FUV. 

\subsection{Additional flags and photometry quality}
\label{subsec:flags}

Instrumental effects, such as the aperture including hotspots or locations with low response in some time bins, can cause extreme changes in brightness during an observation.
These need to be removed before searching for physical variability in time-resolved GALEX photometry using the corresponding flags. 
Below we illustrate variations due to hotspots and low response, as these instrumental effects 
can cause periodic changes in flux that resemble transits, due to the dithering motion.

\begin{figure}

\includegraphics[width=3.5in]{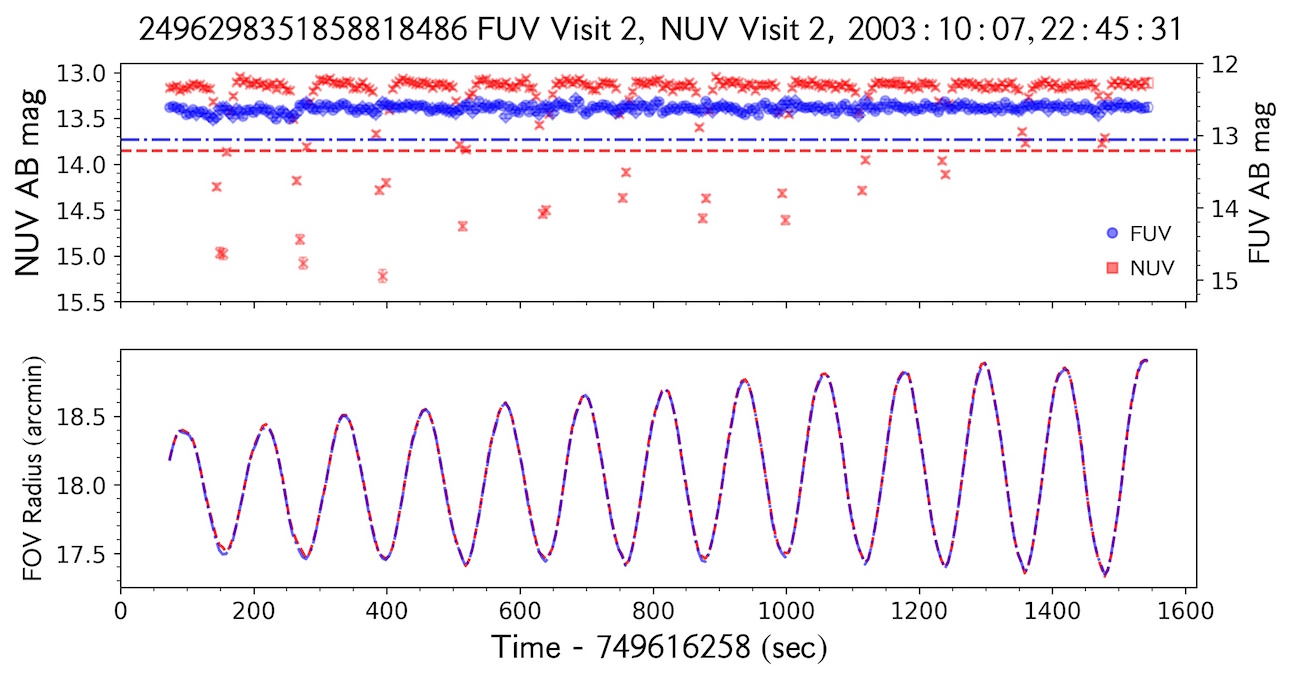}

\caption{An example of a light curve showing variations due to identifiable, known instrumental effects, such as hotspots. 
Hotspot-induced variations frequently correlate with the dither pattern as the source aperture passes over hotspots on the detector. 
Crosses mark hotspot-affected time bins.
Blue dash-dot lines and red dashed lines indicate the FUV and NUV non-linearity cutoffs, respectively.
}
\label{fig:hotspots}
\end{figure}

Figure \ref{fig:hotspots} displays an example of variations caused by hotspots, indicated by crosses. 
Hotspot-generated variability is often periodic and correlates with the dither pattern. 
As sources move in a spiral on the detector during a visit, the source aperture repeatedly crosses regions of the detector affected by hotspots. This
may severely decrease the source count rate, leading to recurring variations $\gtrsim0.5$ mag that mimic transits. Likewise, regions affected by 
low response (i.e. low relative sensitivity) may exhibit large fluctuations in brightness. 
Variations in brightness associated with low response time bins occur in tandem with hotspots more often than not.
Both of these kinds of events are flagged, so the affected data points can be easily removed from the analysis.

In \gphoton, the {\tt detsize} parameter sets the maximum distance of the source from the field center (``{\tt fov\_radius}'' in MCAT and 
GALEX UV Catalogue \citep[GUVCat,][]{bianchi17}) at which photometry is performed. Due to the dithering spiral pattern, which has a typical amplitude 
of the order of $\gtrsim 1$ arcmin, a source within $\sim1'$ of the field's 
edge may get so close to the edge that part of the aperture area or background annulus falls outside the rim, where there are no data, resulting in a drastic magnitude drop. 
The measurement is retained in the {\tt gAperture} output in some cases (see below) but it is flagged. Even when no part of the aperture or background annulus are outside
of the detector, photometry and astrometry may have significantly degraded quality near the edge, and the rim produces severe artifacts 
\citep{bianchi14, bianchi17}. 

The output from {\tt gAperture} provides the $x,y$ position of the source in the detector and the distance from the field center, {\tt fov\_radius}, at any time bin.
The flags ``detector edge'' and ``mask edge'' are set when the average {\tt fov\_radius} for pixels within the aperture exceeds 30 arcmin or is contiguous with 
the detector edge, respectively, and ``bg mask,'' when the background annulus events are contiguous to the detector edge.
To test whether the ``mask edge'' flag is set only when the aperture center is farther than $0.5 \times$ {\tt detsize}
from the center, or also when a portion of the background annulus is beyond the set limit, we varied the {detsize} parameter, 
using {\tt detsize} $=2\times30$' and {\tt detsize} $=2\times36$' for sources with average {\tt fov\_radius} during a visit $\geq30$'. 

Time bins when {\tt fov\_radius} $>33'$ are included in the output both when {\tt detsize}$=2\times33'$ (default value) or $2\times36'$.
The ``mask edge'' or ``bg mask'' flags were never set
in all visits we considered for this test, but the ``detector edge'' flag was always set, as {\tt fov\_radius} $\geq30'$ at all times.

\begin{figure}[h!]
\centering

\includegraphics[width=3.5in]{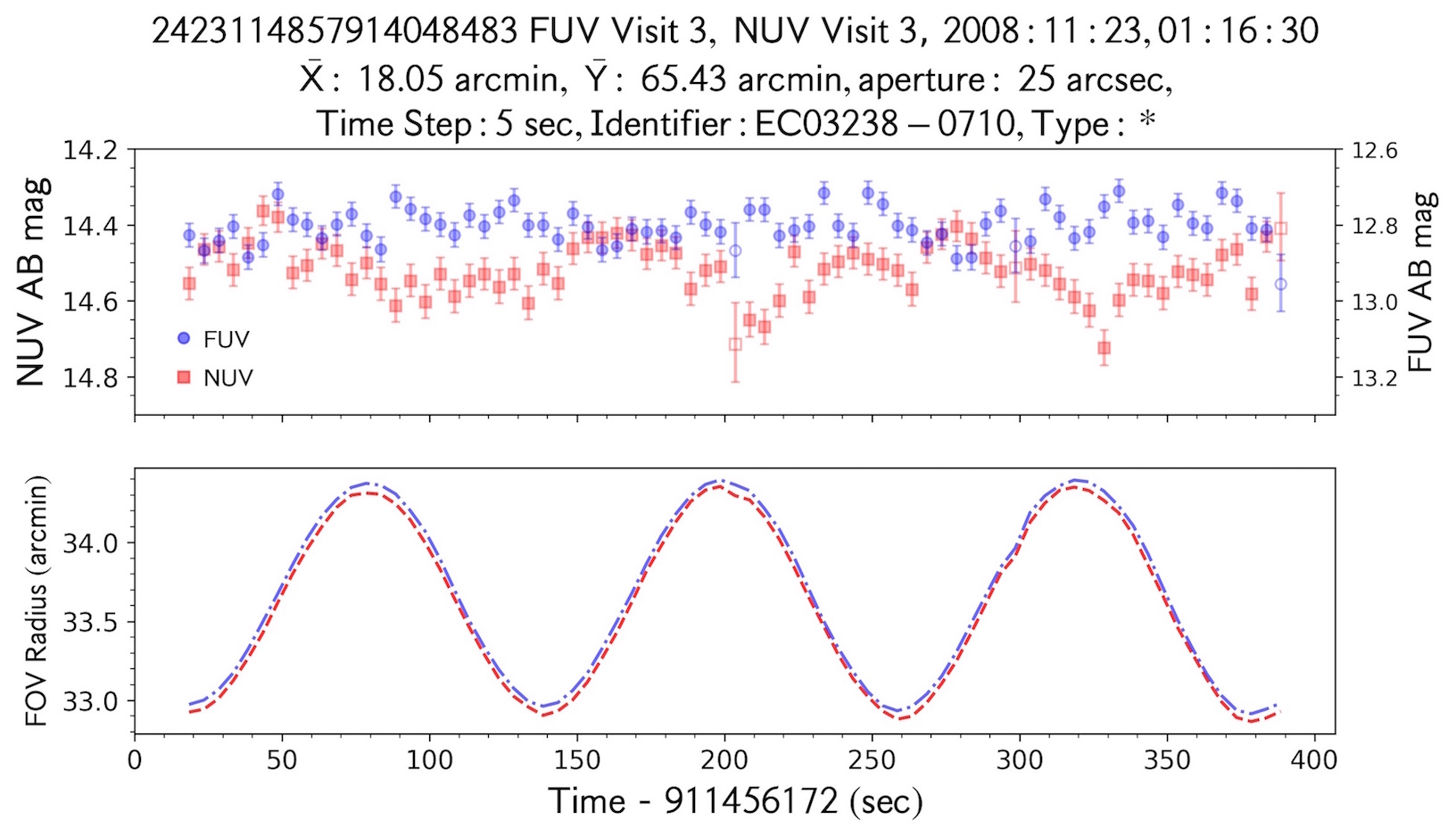}

\caption{Light curve that shows how time bins with {\tt fov\_radius} $> 0.5 \times$ {\tt detsize} are returned by {\tt gAperture} as long 
as there are time bins with {\tt fov\_radius} $\leq 0.5 \times$ {\tt detsize}. 
In this case {\tt detsize} was set to the default value, 1.1$^{\circ}$, and the first and last time bins have {\tt fov\_radius} just under 33 arcmin.
The returned data-points are cleaned by applying a cut in {\tt fov\_radius} for the analysis. 
}
\label{fig:detsize}
\end{figure}

Decreasing {\tt detsize} to $2\times30'$ 
removes time bins during which the target {\tt fov\_radius} is $>0.5 \times$ {\tt detsize}.
If there are time bins when {\tt fov\_radius} $\leq 0.5 \times$ {\tt detsize}, the first and last bins satisfying this criterion
are returned in the {\tt gAperture} output, as well as all time bins in between, even if the {\tt fov\_radius} exceeds $0.5 \times$ {\tt detsize} in between these times. 
Figure \ref{fig:detsize} demonstrates this effect for the default {\tt detsize} value of 1.1$^{\circ}$: 
even though the majority of the {\tt fov\_radius} values throughout the 
visit exceed 33 arcmin, data are computed and included in the output as the first and last time bins during the visit $\leq 0.5 \times$ {\tt detsize}.
The ``mask edge'' or ``bg mask'' flags  were not set for bins when {\tt fov\_radius} exceeds {\tt detsize}, when {\tt detsize} is set to 30'. 
Therefore, cuts were applied post-facto using the {\tt fov\_radius} value for each measurement. 

\begin{figure*}
\centering
\includegraphics[width=6.5in]{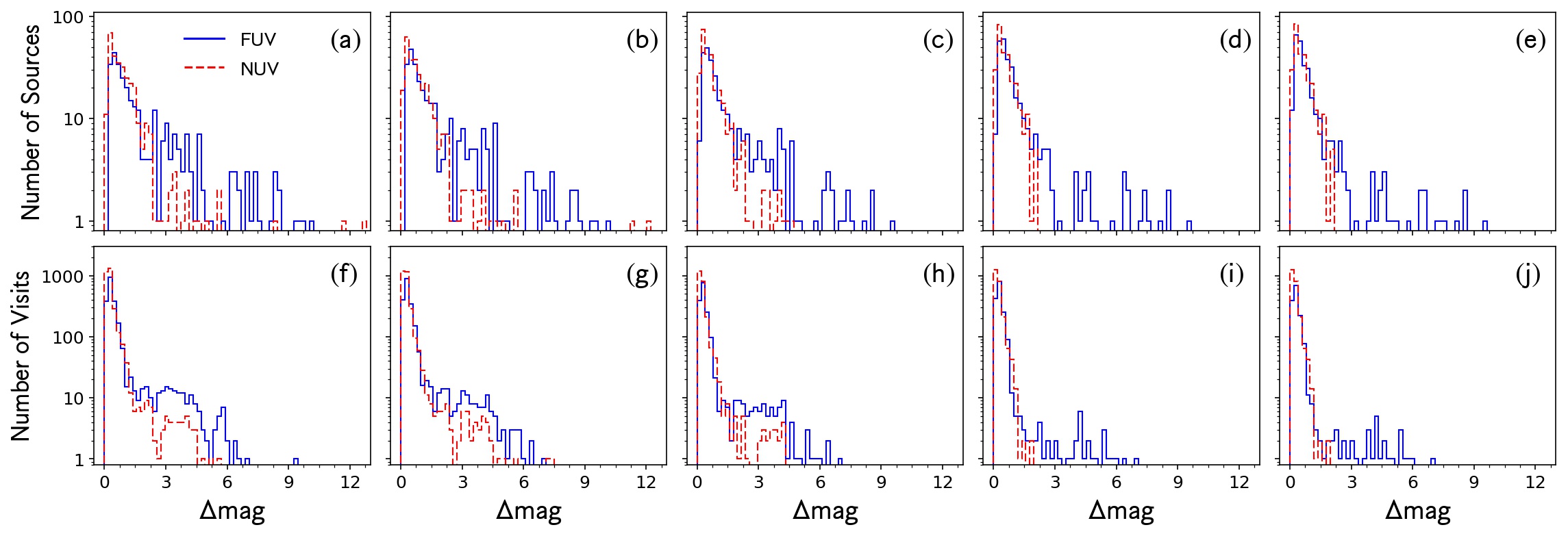}
\caption{Top row: Histogram of largest variation in magnitude, $\Delta \textrm{mag}$, {\bf (a)} for all sources in our sample, and
after successively removing:
{\bf (b)} time bins with exposure time $< 75\%$ of the step size; 
{\bf (c)}: time bins with detector radius $>30$ arcmin; 
{\bf (d)}: time bins with hotspot flags;
{\bf (e)}: time bins with low response flags.
Bottom row: Same as top row but for visits. The last plots ({\bf (e)} and {\bf (j)}) are the ``clean sample.''
The vast majority of visits have significant variations that are not due to the instrumental effects we eliminate in this plot. 
}
\label{fig:max_mag_var}
\end{figure*}

\begin{figure}
\centering
\includegraphics[width=2.5in]{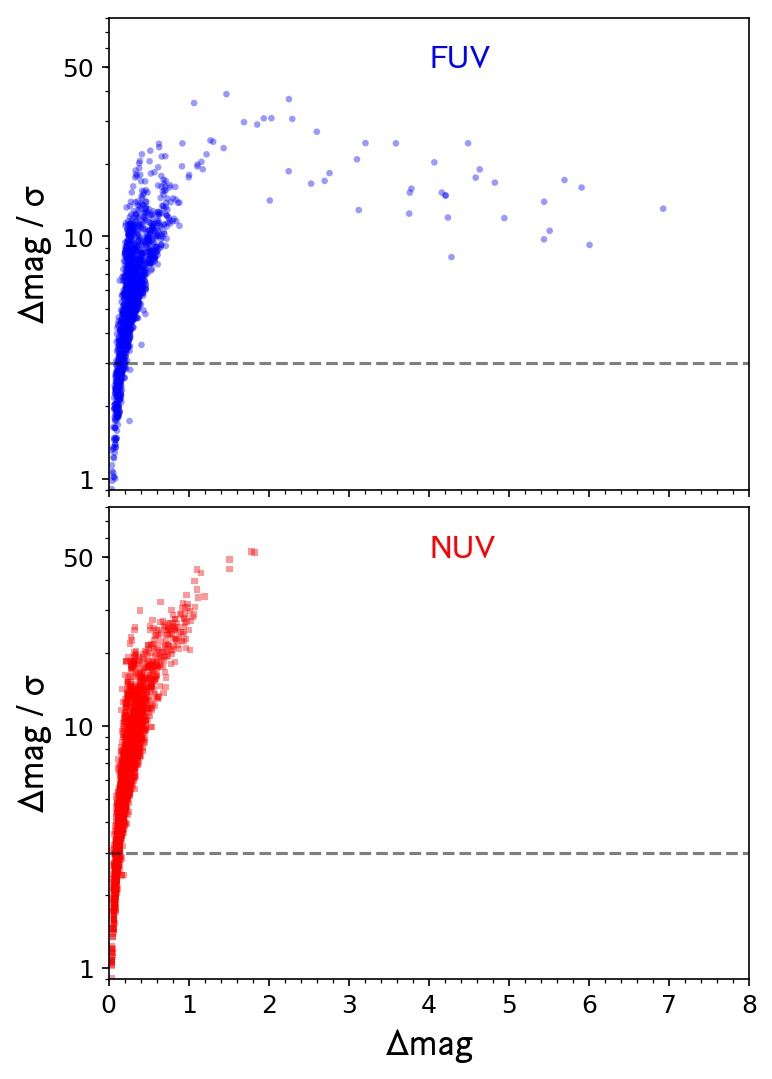}
\caption{Maximum variation versus significance of the variation per visit separated by band for the ``clean sample.''
The vast majority of visits have variations less than half a magnitude, with FUV exhibiting most of the 
large variations, as also seen in Figure \ref{fig:max_mag_var}. 
The dashed line indicates visits with maximum variations $=3 \sigma$.}
\label{fig:max_mag_var_visits}
\end{figure}

\section{Analysis. Searching for Variability}
\label{sec:analysis}


Variations can be non-periodic, such as flares or transient phenomena, or periodic, and these may or may not be detectable depending on
the serendipitous coverage and cadence of the data with respect to the period. In order to examine our large sample of 
over 5000 visits, we first identify visits where significant variations occur, inspect them, and run Fourier analysis and period searches.
In this section we identify visits with significant variations. 

We searched for variability in our sample by computing within each visit the maximum range in brightness, $\Delta \textrm{mag}$, and its error, $\sigma$. 
The significance of this variation is simply $\Delta \textrm{mag} \ / \ \sigma$.
We computed the maximum variation in brightness as the difference between the average of the three highest magnitudes and the average 
of the three lowest magnitudes in each visit among all time bins satisfying the criteria outlined in Section \ref{subsec:flags}.
The error of this maximum variation is the sum in quadrature of the average error on the faintest magnitudes 
and the average error on the brightest magnitudes. 
For reference, the typical error on a time bin in our clean sample is 0.03 mag. 
We also searched for variability over long time scales by computing the maximum variation between all measurements of each source in our sample.
We calculate the largest variation in brightness and its error the same way as 
is done for each visit, except we consider measurements across all visits.

In Figure \ref{fig:max_mag_var} we show the distribution of maximum variation in brightness, 
$\Delta \textrm{mag}$, for all visits and sources in our sample. In the top row, from left to right, we show histograms of $\Delta \textrm{mag}$ 
for all measurements (not culled), and after excising, in succession,
measurements with exposure time $<75\%$ of the time bin, {\tt fov\_radius} $> 30$ arcmin, hotspot flags, and low response flags (``clean sample''). 
In the bottom row we show the same, but for visits.
Figure \ref{fig:max_mag_var} shows that our culling criteria to define a ``clean'' analysis sample (Section 2.4) 
eliminate many of the largest variations ($\Delta \textrm{mag} > 1$), which are spurious. The vast majority of both visit- and source-level maximum variations 
have values of $\Delta \textrm{mag}\sim0.2$, however a non-negligible number of very large ($\sim$ few mag) variations persists.
Figure \ref{fig:max_mag_var_visits} shows the significance of these variations, $\Delta \textrm{mag} \ / \ \sigma$, against $\Delta \textrm{mag}$ in the .
The vast majority of visits have significant ($\Delta \textrm{mag} \ / \ \sigma \geq 3$) variations $< 0.5$ mag. 

Events with {\tt fov\_radius} beyond 30 arcmin and hotspot flag set contribute the most to the occurrence of extreme variability (138 and 130 visits, respectively), 
when compared to effective exposure time $\leq 75\%$ of the time bin
and low response (25 and 3 visits, respectively). However, there exist a few dozen visits with extreme variations not due 
to any instrumental effects reported thus far, mainly in the FUV. These variations correspond to an artifact variation we report in Section \ref{subsec:rise},
a rapid rise in FUV brightness that generally occurs when visits commence.

\begin{figure}
\centering
\includegraphics[width=3.6in]{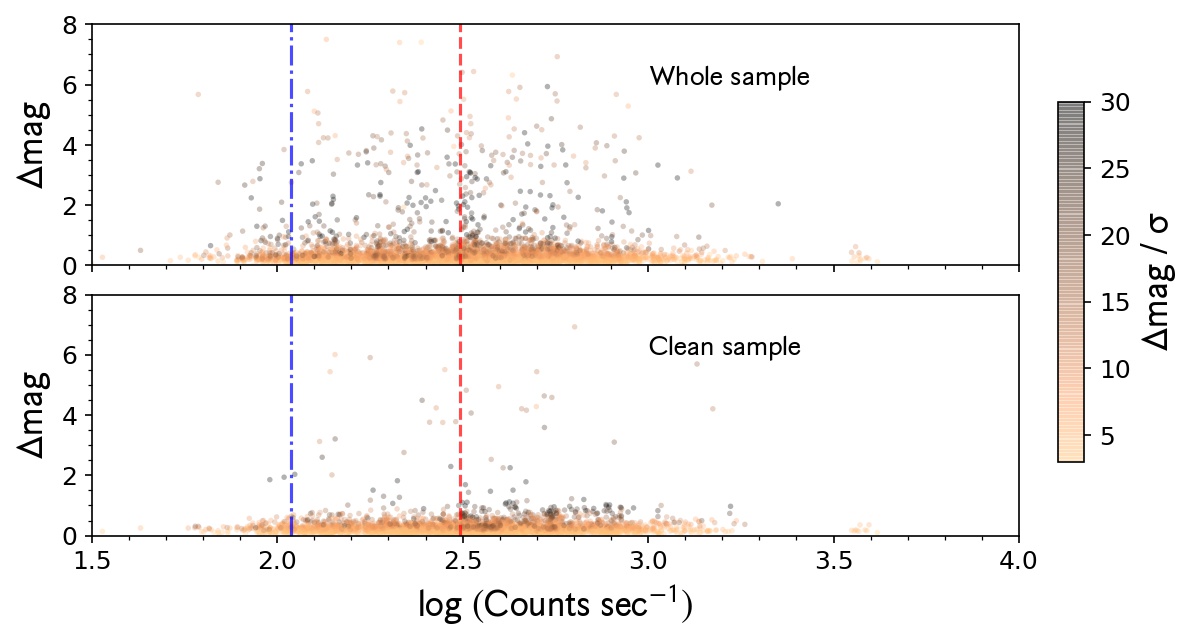}
\caption{Maximum variation within each visit against average, background-subtracted count rate per visit colored by the significance of the variation. 
The top panel includes all data points. The bottom panel excludes data affected by short exposure times, hotspots, low response and proximity to the detector edge
(``clean sample''). The blue dash-dot and red dashed lines indicate the 10\% non-linearity cutoff in the FUV and NUV, respectively. }
\label{fig:max_mag_var_cps}
\end{figure}

Figures \ref{fig:max_mag_var} and \ref{fig:max_mag_var_visits} show that 
instrumental effects such as hotspots, low response, short exposure time, or proximity to the detector edge do not account 
for all of the variability in our bright GALEX sample. Two other causes for spurious variability mentioned by M16 are 
count rate above the 10\% non-linearity cutoff, and {\tt fov\_radius} within the visit. We examine these effects in our sample. 
Figure \ref{fig:max_mag_var_cps} shows $\Delta \textrm{mag}$ versus average, background-subtracted count rate for all visits (top panel) 
and clean sample (bottom panel).
Aside from a few significant variations with $\Delta \textrm{mag} > 1$, the distribution
of $\Delta \textrm{mag}$ as a function of count rate in the bottom panel is uniform across an order of magnitude 
and does not significantly differ above the non-linearity limit. 

\section{Artifact-induced variability}
\label{sec:artifacts}

When we began to analyze the time-resolved photometry of our sample, a number of variations became readily apparent, which we suspected to be instrumental
effects given their unrealistically high occurrence rate, as well as their characteristics. 
We describe below the five major types of artifact variations found in our sample,
and investigate the causes inducing such instabilities. None of these effects was previously reported or discovered, therefore we 
examine the frequency of their occurrence and look for possible correlations with instrumental parameters.

We select visits with duration longer than 200 sec, which leaves 3959 light curves.
In an attempt to isolate changes in brightness which are just a result of artifacts and reduce noise, we
bin light curves to 10 sec resolution and sigma-clip time bins greater than 2.5 standard deviations from the mean. After this cut, we calculate
$\Delta\rm{mag}$ and its uncertainty in the same fashion as in Section \ref{sec:analysis}.
We used the maximum variation $\Delta \textrm{mag}$ calculated for each visit 
(Section \ref{sec:analysis}) to select visits with significant variations. 
We run Fourier analysis of the light curve of each visit.


\subsection{Short period ($P \sim 120$ sec) quasi-sinusoidal variations}
\label{subsec:triangles}

\begin{figure*}
\centering

\includegraphics[width=1.15in]{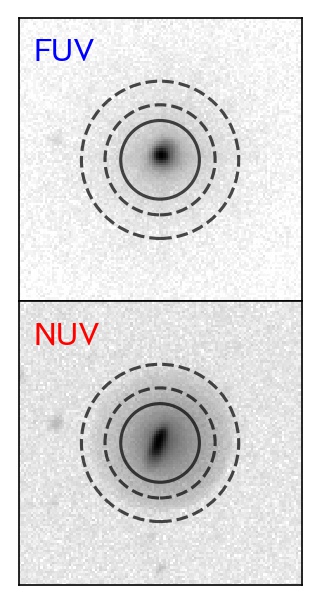}\includegraphics[width=5.in]{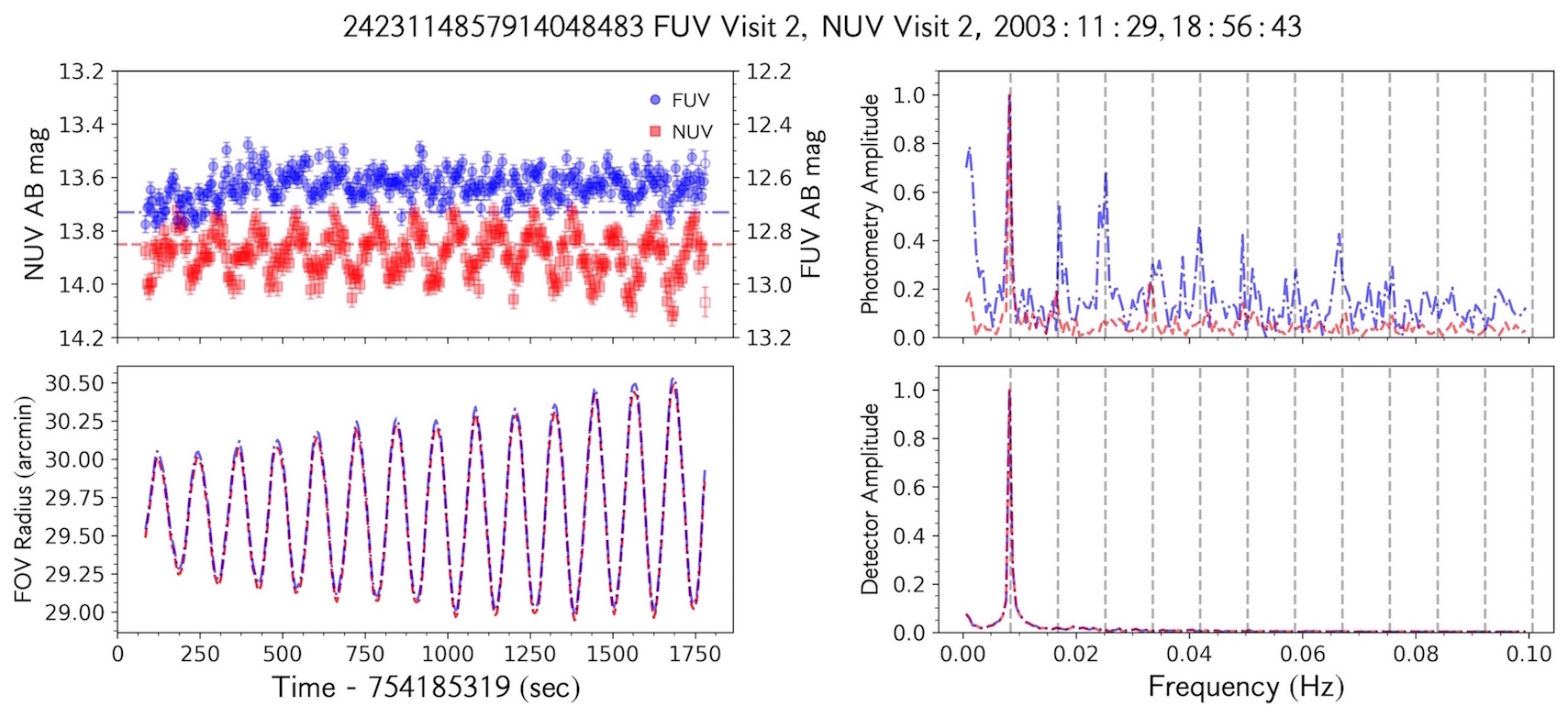}
\includegraphics[width=0.7in]{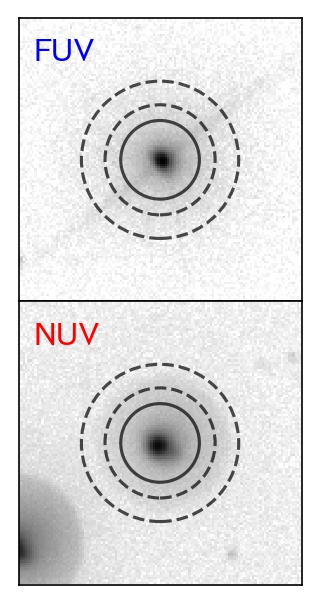}\includegraphics[width=2.5in]{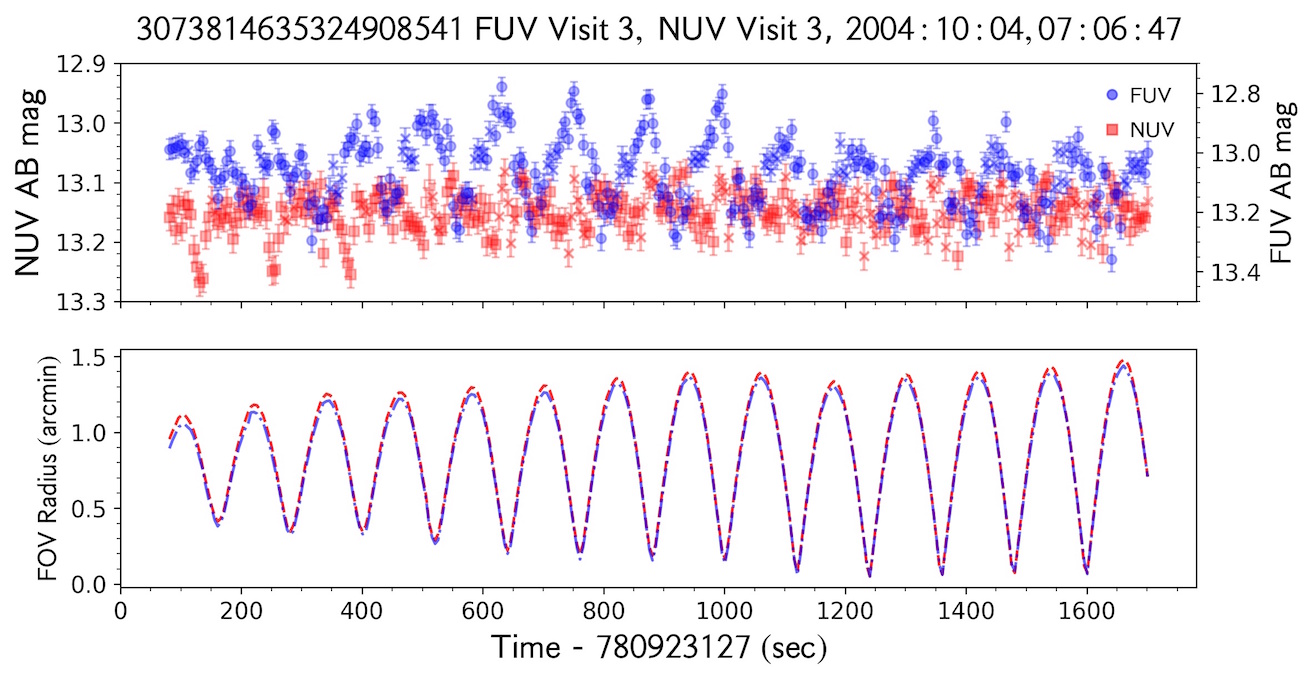} \includegraphics[width=0.7in]{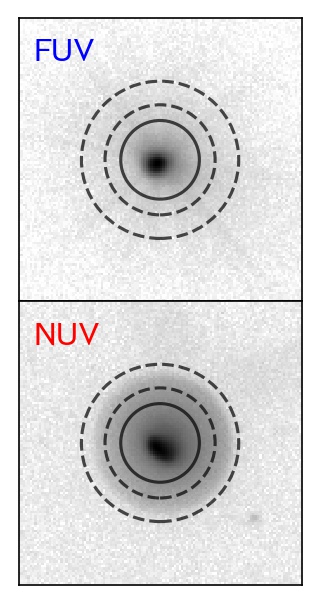}\includegraphics[width=2.5in]{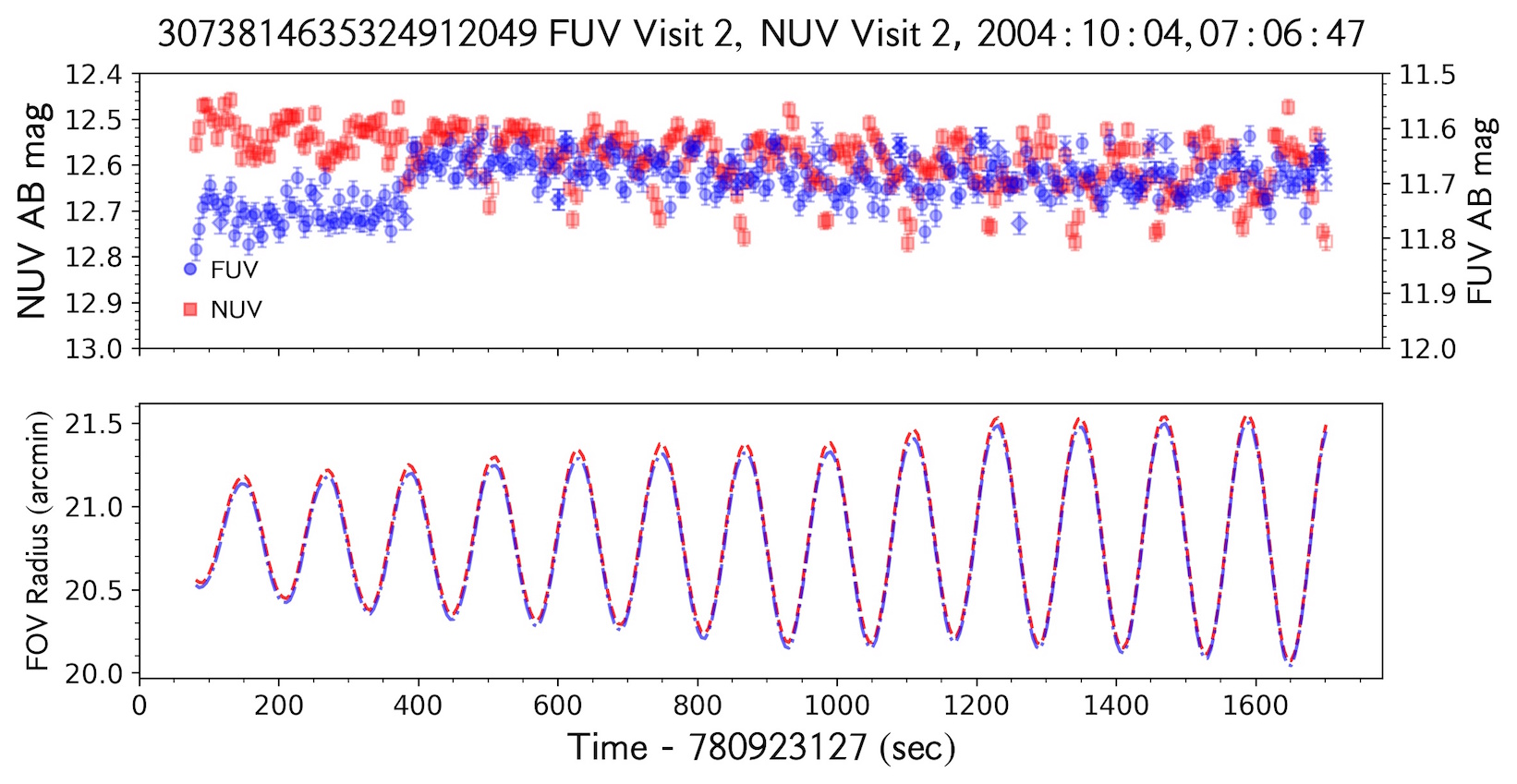}
\includegraphics[width=2.1in]{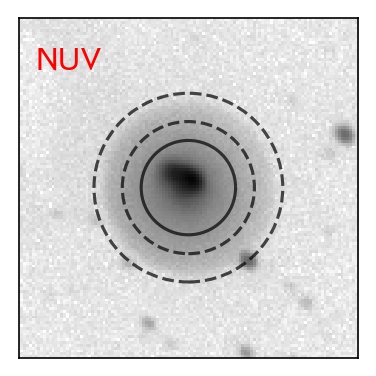}\includegraphics[width=5in]{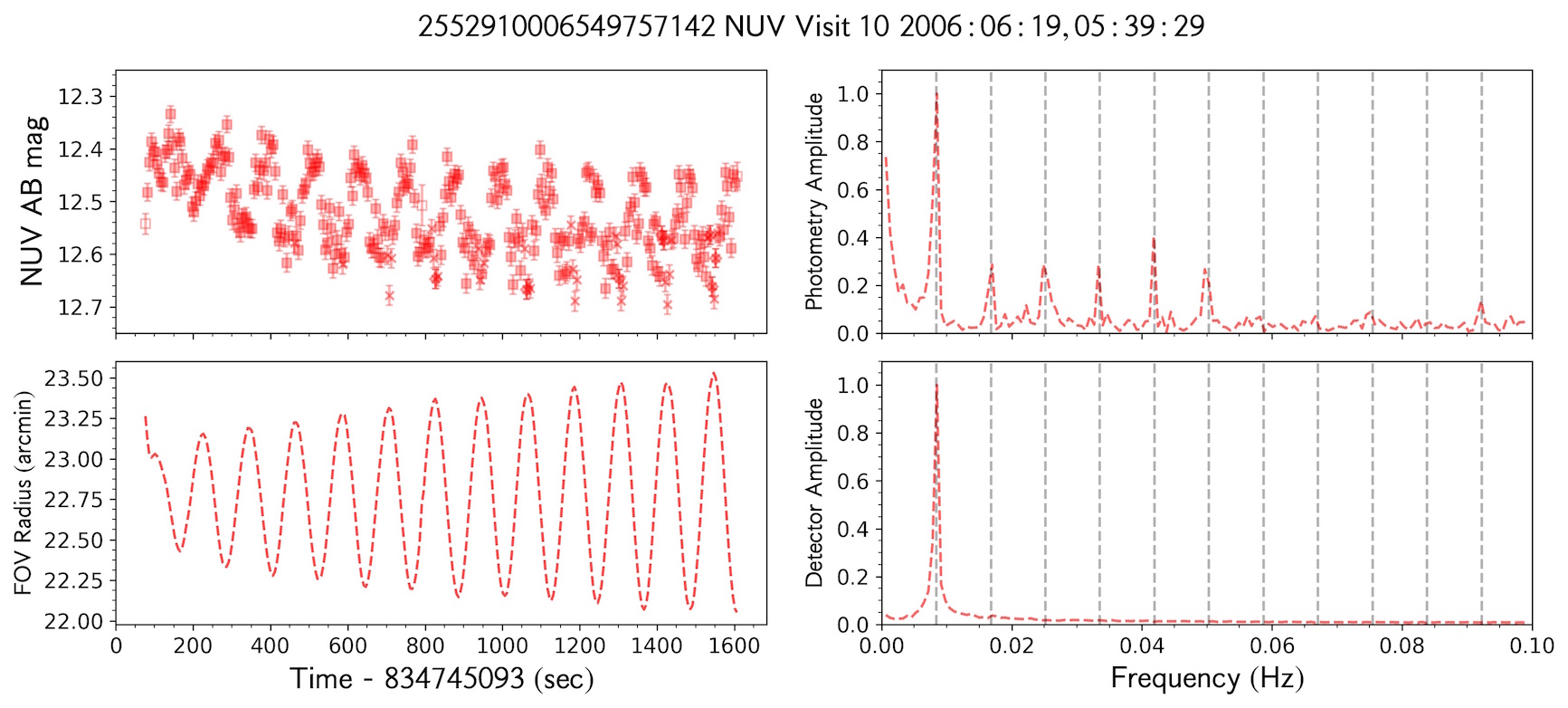}

\caption{Examples of light curves exhibiting ``triangle-wave,'' quasi-sinusoidal variations, labeled by the object ID and observation date. 
Top rows: FUV and/or NUV light curves in the top left plot, dither pattern in the bottom left, 
FT of the FUV and/or NUV photometry in the top right, and FT of the dither pattern in the bottom right. 
Quasi-sinusoidal variations do not necessarily occur in both bands.
Middle row: light curves for two sources from the same observation, in different regions of the detector.
These light curves demonstrate that the triangle-wave variation is not a detector-wide artifact. 
Bottom rows: with same layout as top row, a light curve showing both triangle-wave and slope variation. 
All FTs are normalized by the peak amplitude. 
Hollow markers, cross markers, and diamond markers
indicate short exposure time-, hotspot-, and low response-affected time bins, respectively, which are excluded from the analysis.
Blue dash-dot lines and red dashed lines indicate the FUV and NUV non-linearity cutoffs, respectively.
To the left of each light curve is a 3' by 3' finding chart centered on the source in each band. 
The inner solid circle is the 25'' aperture and the dashed circles are the 35" and 50" boundaries of the background annulus.
}
\label{fig:triangles}
\end{figure*}

The first striking result of our analysis was the detection of a strictly periodic, 
short period ($P$ between 100 and 120 sec) almost sinusoidal (``triangle-wave'')
variation, of which we show examples in Figure \ref{fig:triangles}.  

As the three examples illustrate, such variations do not necessarily occur in all observations of the same source, and not always in both bands; sometimes they 
appear in only one band, sometimes they are correlated in both bands but may also not be correlated. We recall that GALEX FUV and NUV fluxes are recorded 
simultaneously in two separate detectors, through a dichroic beam splitter; the occasional unmatched behavior in FUV and NUV was also a first indication of a 
possible instrumental cause. However, the target shown in Figure \ref{fig:triangles} 
is a cataclysmic variable (CV), as are other bright sources in our sample (because of the  color selection FUV-NUV $<$0), 
therefore this type of variation and even differences at different wavelengths are not unexpected; the curve shape and short period are not unreasonable 
for hot stellar pulsators. A stronger indication that the variation may have instrumental origin rather then being a physical pulsation came for the very high detection rate of 
similar variations across the sample, their period falling in a narrow range. In other space instrumentation,  response variations are related for example to the detector 
temperature, which may be influenced by possible nearby heaters, but after consulting with the instrument experts (P. Morrissey, priv. comm) this was ruled out for GALEX.  
Final proof of the instrumental origin of such ``fake pulsations'' came from plotting the target position in the detector during the exposure:
the distance from the target center ({\tt fov\_radius}) is shown in Figure \ref{fig:triangles} in the bottom plots of the first and third rows, 
and the dithering spiral pattern was found to be synchronized with the photometric variation in all cases when this occurs. 

\begin{figure}
\centering

\includegraphics[width=0.75in]{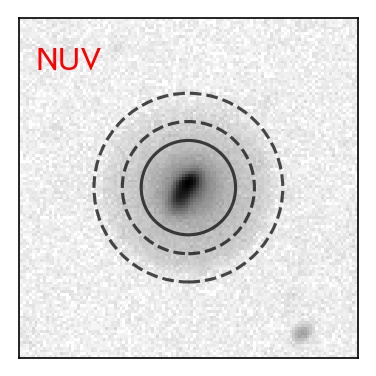}\includegraphics[width=2.75in]{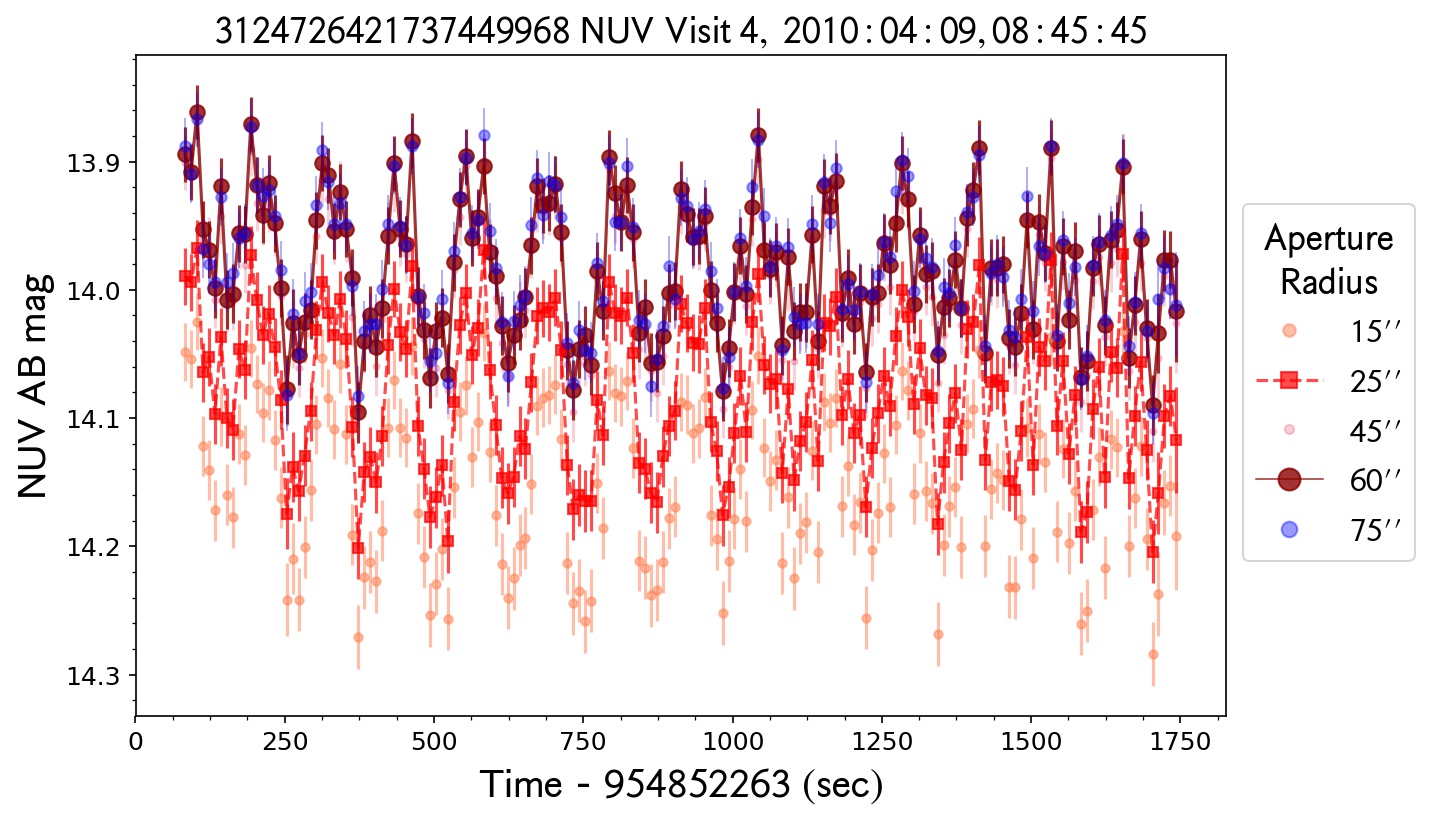}

\caption{Light curve with different photometric apertures for a source 
exhibiting triangle-wave variations, illustrating that variability persists in apertures
with radii larger than the typical dithering amplitude. For clarity, we rebin time elements to 10 sec and emphasize 60'' (dark red circles), 
the standard dither amplitude, and the largest aperture considered, 75'' (light blue circles). 
}
\label{fig:var_aper}
\end{figure}

One way that a dithering motion may cause variation of the flux measurements is if the image reconstruction and corresponding centering of the photometric aperture  
performed by \gphoton, which integrates photon events over the specified time bin, were not adequately following the spacecraft attitude motion. In other words, if the 
aperture centering was not precisely compensating the dithering spiral, part of the flux would wander in and out of the photometric aperture. While the GALEX PSF is $\sim$ 
4.2'' (FUV) / 5.3'' (NUV), there is a considerable ``skirt'' around the central peak, especially in NUV and particularly evident for bright sources. 
If aperture centering were the cause for the variations, increasing the 
aperture radius to values comparable to the amplitude of the uncompensated motion would reduce the variation. Given that the total dithering amplitude is of the order of 
1 arcmin, we performed tests increasing the aperture radius up to 75'' for visits exhibiting quasi-sinusoidal variations, to test whether the variability persists
at larger apertures. An example of this test is shown in Figure \ref{fig:var_aper}. The short-period variation still appears, its amplitude not decreasing, confirming 
the accuracy of \gphoton's  astrometry reconstruction and ruling out this cause of this variability. Additional confirmation was provided by examination of the 
background counts, which, if the star were wandering out of the aperture, would be affected in the opposite way to the source. 
For isolated sources (i.e. no other source contaminates either the aperture or the background annulus), we have examined the background count-rate and found no 
fluctuations in brightness during visits where the source exhibits periodic variations related to the dithering motion. 

To test whether triangle-wave periodic variations arise from detector-wide effects, such as voltage or temperature fluctuations, we examined other sources in the field
during the same observations of a source that exhibits this artifact. We found that other sources in the field do not necessarily display this artifact. 
The middle two plots of Figure \ref{fig:triangles} exemplify this behavior in light curves of two different sources observed in the same visit.  
Both sources display periodic variations synchronized with the dither, but with very different amplitude and shape between them and between the two detectors. 
One source (3073814635324912049) also displays a non-periodic flux ``jump'' (described in Section \ref{subsec:jump} below) in FUV of $\sim$0.1~mag, and 
increasing amplitude of both flux variation and {\tt fov\_radius} pattern in NUV. 
The other source (3073814635324908541) exhibits marked triangle-wave variation in FUV, whose amplitude varies but not in a way correlated to the dither amplitude. 
We also investigated whether sources that show triangle-wave variations in one visit do so in all visits, and found that this is not the case for any source in our sample.

The majority of our sample lies in the non-linear regime. While a 14th magnitude point source corresponds to 
85 counts $\textrm{sec}^{-1}$ in FUV and 270 counts $\textrm{sec}^{-1}$ in NUV (below the respective 10\% rolloff levels), the point spread function (PSF)
core may still be distorted at these count rates. Most likely there is a large ($>10\%$) effect in the PSF core and a smaller effect
in the wings. The white dwarf calibration star LDS749B, for example, produced 140 counts $\textrm{sec}^{-1}$ in NUV and required a 20\%
correction at the edge of the field due to saturation effects. Distortion in the PSF varies as the source moves around the detector
due to the changing ability of the detector to source sufficient current (gain) at different locations. The count rate corresponding to a 10\% rolloff
will fluctuate around the detector due to the geometry of the microchannel plates, in particular the way the plates are clamped at the edges (we thank the referee 
for providing the above detail).

Given the obvious correlation of the periodic variation with dithering, but its non-ubiquitous occurrence, and the severe complication that it poses for analysis of stellar pulsations, 
which may have similar (or smaller) amplitude and periodicities, and for which the \gphoton \ database enables for the first time a comprehensive search, 
we tried to establish when the dithering pattern causes such variations (in each detector) and when it does not.   
To answer this question, we identified light curves that exhibit such triangle-wave variability and no other types of spurious variation (discussed in the following sections). 
We first selected light curves with $\Delta \textrm{mag} \leq 0.5$ mag in the .
We measured the peak frequencies from the FT spectrum of the source {\tt fov\_radius}, 
which gives the exact period of the dithering motion in the specific light curve, then searched for peaks of this frequency (and its first few harmonics) in the FT  of the light curve, 
and measured the significance of such FT peaks with respect to the noise level at nearby frequencies.
After choosing light curves with significant FT peaks (amplitude $> 3\times$ the FT noise level) in their photometry
with frequencies within 20\% of the FT peak frequency in the dither pattern, we visually inspected these light curves to confirm the presence of artifact variations. 

For the light curves that exhibit solely triangle-wave variations due to the dithering pattern, we plot
$\Delta \textrm{mag}$ versus  the Julian Date, average {\tt fov\_radius}, and average source (background-corrected) count rate (Figure \ref{fig:jpc_labels}).  
The top-most plot indicates that triangle-wave flux variations occur in GALEX data from the beginning, 
and persisted for the entire duration of the mission, with no clear dependence of $\Delta \textrm{mag}$
on the date of the observation. Therefore, they cannot be ascribed to periods when the detectors suffered occasional problems, or to secular decay. 

In the middle panel, top row of Figure \ref{fig:jpc_labels}, we plot $\Delta \textrm{mag}$ against the mean {\tt fov\_radius} during the visit. 
Most detected triangle-wave variations occur
at ${\tt fov\_radius} > 10$ arcmin, and $\Delta \textrm{mag}$ gently increases with {\tt fov\_radius}, though the majority of $\Delta \textrm{mag}$ is concentrated around 0.2 mag.
We plot the mean X,Y positions of all sources displaying triangle-wave variations, as well as all visits in our sample, in Figure \ref{fig:triangle_quad} and note the fraction 
of occurrence of the artifact over the total number of visits in each band in five radial bins, each 7 arcmin wide in {\tt fov\_radius}.
More concisely, we plot the artifact fraction for all artifacts in Figure \ref{fig:quad_artifact} as a function of {\tt fov\_radius}, separated by quadrant and band.
We summarize these fractions in Table \ref{tab:artifact} for the first four artifacts discussed in this section.
Uncertainties on the artifact fraction are assumed to be Poisson, i.e. the square root 
of the number of artifacts divided by the total number of visits, and are typically less than 5\%.
Triangle-wave artifacts occur in $\sim5\%$ of FUV visits and $\sim10\%$ of NUV visits 
at ${\tt fov\_radius} < 15$ arcmin and 25\% to 50\% of NUV visits ${\tt fov\_radius}> 15$ arcmin.
These figures demonstrate that, throughout the detector, triangle-wave artifacts occur more in NUV and towards the detector rim. 
Beyond 25 arcmin, triangle-wave variations account for a quarter of visits, except for the 
third quadrant, which has artifact fraction $<15\%$ at all {\tt fov\_radius}. 

All the above tests lead to the conclusion that the dithering-synchronized variability is due to small-scale response variations, which are not accounted 
for in the general instrument calibration. Indeed, the purpose of the dithering pattern was to smooth out pixel-to-pixel variations, and the effect of these local 
response variations had not been previously quantified. 
The fraction and $\Delta\rm{mag}$ of the artifact variation increase with {\tt fov\_radius} out to 35 arcmin.

Amplitudes of the triangle-wave variation seem to mildly anticorrelate with count rate, but we recall that most of our sample is brighter than the non-linearity onset flux level, 
as the aim of this work is to examine the potential for time-domain studies at short timescales, and the highest count rates allow analysis of the smallest
variations even with short integrations. 
The rightmost top panel of Figure \ref{fig:jpc_labels} shows, as also illustrated in our examples, that such variation occurs also in the linear regime. 


\subsection{``Jump'' Variations}
\label{subsec:jump}

\begin{figure*}
\centering

\includegraphics[width=1.15in]{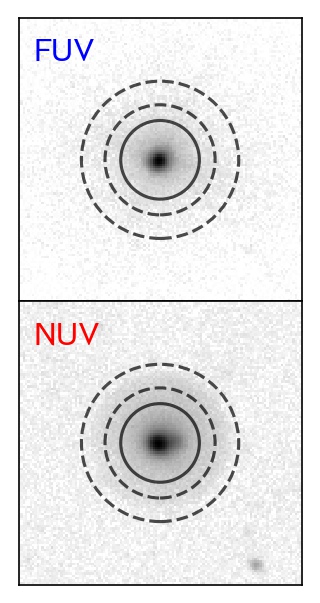}\includegraphics[width=5.in]{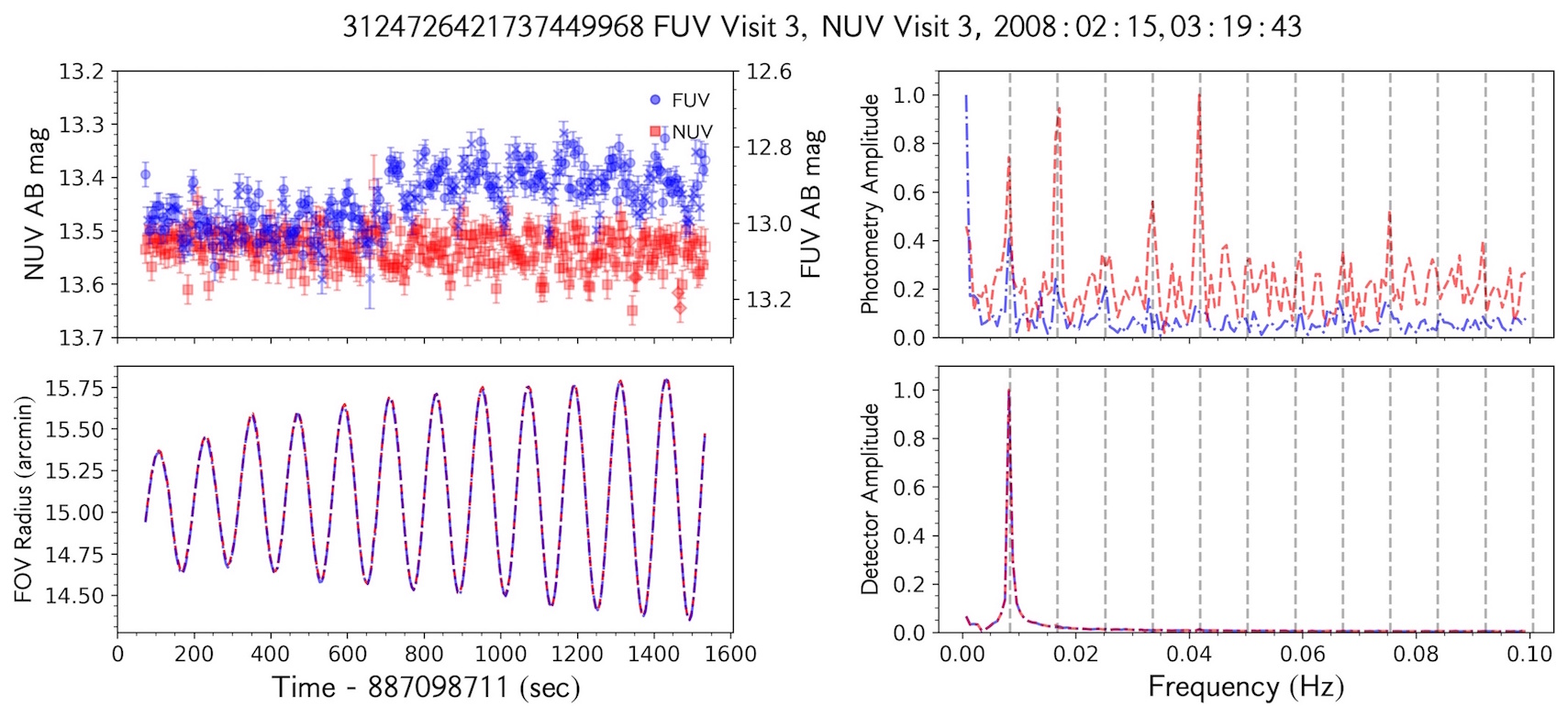}
\includegraphics[width=1.15in]{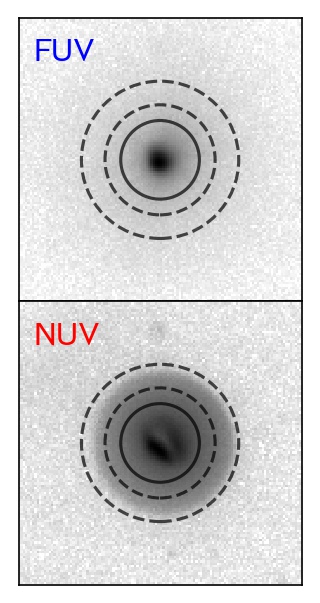}\includegraphics[width=5.in]{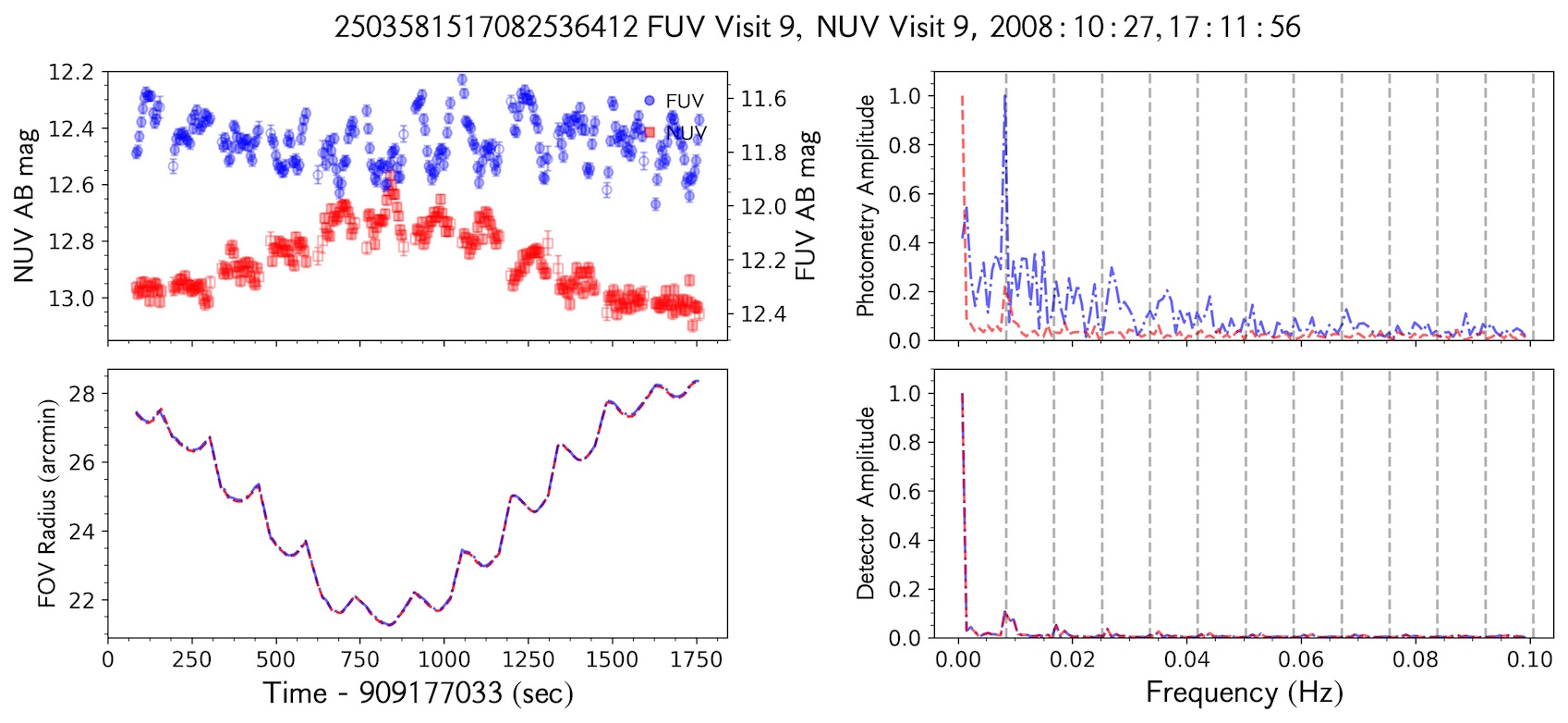}
\includegraphics[width=1.15in]{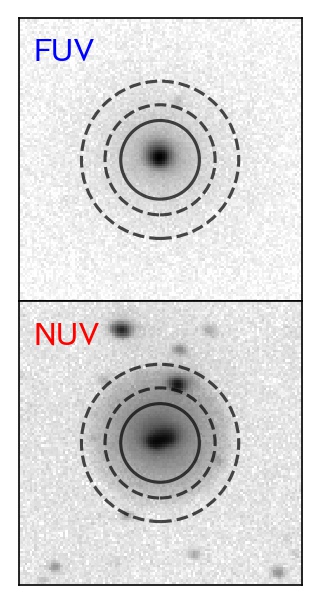}\includegraphics[width=5.in]{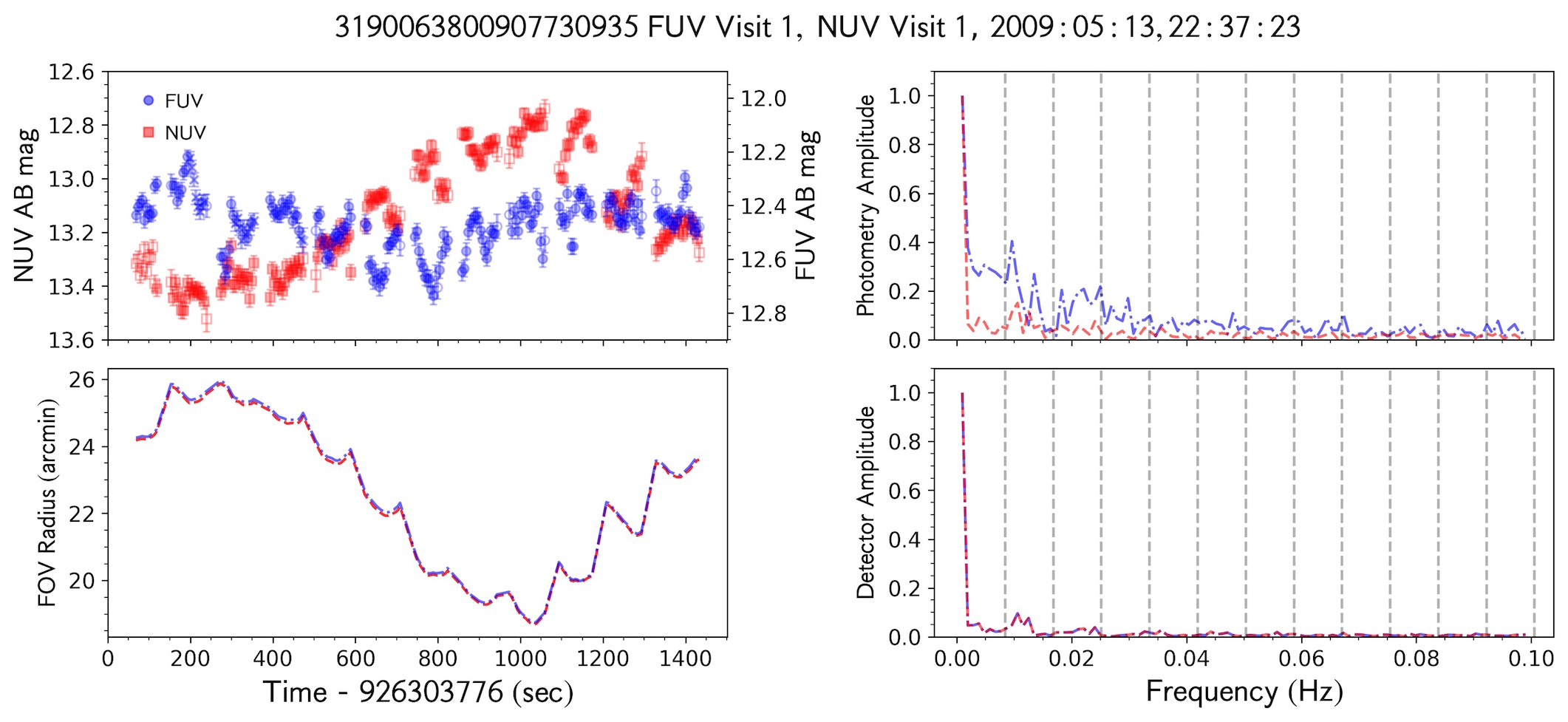}

\caption{Light curve and dither position plotted as in Figure \ref{fig:triangles}, showing examples of 
a ``jump'' artifact (top panels) and two ``sagging'' cases (middle and bottom panels). 
As with triangle-wave variations, jumps often do not happen in both bands in the same visit. 
``Sagging'' variations may differ in sign and in amplitude between the two bands, such as in the bottom panel, 
where the FUV light curve varies by $\sim0.5$ mag, while the NUV light curve varies by $\sim0.8$ mag.
}
\label{fig:jumps}
\end{figure*}

Another type of artifact variation we detected is a smooth, one-time increase in brightness that takes $\sim100$ sec to transition and typically has amplitude $\sim0.2$ mag. We provide 
example light curves in the middle panels of Figure \ref{fig:triangles} and the top panel of Figure \ref{fig:jumps}. 
``Jumps'' happen almost exclusively in FUV light curves, and often do not appear in both bands during the same visit.
Unlike the triangle-wave variation, jumps do not correlate with the dither pattern and typically occur within the first 500 - 800 sec of the visit.  

As for the previous case, we examine whether jumps correlate with visit parameters. 
Jumps are seen since the beginning of the mission, though there is a noticeable dearth of occurrences $\sim700$ days after GALEX launched in our sample (Figure \ref{fig:jpc_labels}).
This time coincides with a major FUV anomaly recovery in March - August 2005, during which the FUV detector voltage was cycled on and off and could 
result in large changes in brightness, depending on the voltage level.
We note that in the period when we do not see ``jumps'' in our sample, we do see another type of artifact, described in Section \ref{subsec:rise}.

We investigated whether the jump artifact is a detector-wide variation by examining other sources in the field during 
some of the visits where the jump is observed, with negative results.
As an example, the source in the middle row, right panels in Figure \ref{fig:triangles} exhibits a jump in the FUV light curve,
but the source in the left panels, observed at the same time, does not. 
``Jumps'' appear to be a local effect and do not occur at specific times during the visit.
Detector temperature and voltage readings from the spacecraft state files show no correlation with the occurrence of jumps.


Jump artifacts exhibit broad dispersion in occurrence rate as a function of {\tt fov\_radius} (see second panel from top in Figure \ref{fig:jpc_labels}),
and with respect to quadrant on the detector (Figure \ref{fig:quad_artifact}).
Jumps are overwhelmingly a FUV artifact, with all NUV artifact fractions less than 3\%.
FUV occurrence rate ranges from 15 to 30\% for ${\tt fov\_radius} < 30$ arcmin and drop to $\lesssim10\%$ beyond 30 arcmin. 
We see little dependence on $\Delta \textrm{mag}$ with {\tt fov\_radius} (aside from the fact that more jumps occur at ${\tt fov\_radius} > 10$ arcmin)
and a wide spread in $\Delta \textrm{mag}$ at low count rates. 

\subsection{``Slope'' Variations}
\label{subsec:slope}

We occasionally observe a smooth, non-periodic decline in brightness, typically by $\sim0.2$ mag over a span of $\sim500$ sec before the flux stabilizes. An example of this 
variation is shown in the bottom panel of Figure \ref{fig:triangles}. Nearly all ``slope'' artifacts occur in the NUV. Like jumps, they are not correlated with the dither pattern 
and do not appear in both bands in the same visit. Akin to jumps, slopes can occur 
simultaneously with triangle-wave variations, as seen in the bottom panel of Figure \ref{fig:triangles}.

We tested whether the slope artifact emerges from a detector-wide effect by examining other sources on the detector 
observed at the same time. Other sources observed at the same do not always show slopes variations, ruling out this artifact as a detector-wide cause. 

In Figure \ref{fig:jpc_labels} we plot the amplitude of the artifact $\Delta \textrm{mag}$ versus other parameters. 
``Slope'' artifacts occur throughout the length of the mission,
mostly at {\tt fov\_radius} $20 < R < 30$ arcmin. Amplitudes show no correlation with {\tt fov\_radius} 
or source count rate, and slopes are also seen for count rates within the linear regime. 
NUV artifact fractions per quadrant display little spread, lying within statistical uncertainties at all {\tt fov\_radius}. Fractions gently rise from $\lesssim 5\%$
at a {\tt fov\_radius} of 10 arcmin, to $\sim12\%$ at the edge of the detector. 
A future work will address a fainter sample and the comparison may provide additional clues, as worse high-count rate performance is expected at large {\tt fov\_radius} values.

\subsection{``Sagging'' Across the Visit}
\label{subsec:sagging}

Another significant artifact we found consists of large changes in brightness ($0.3 - 0.5$ mag in most cases), 
resembling a smooth ``heaving'' or ``sagging'', in response to a larger-than-normal dither pattern that completes one cycle in an observation. 
We refer to this artifact as ``sagging'' for brevity. Examples are shown in the middle and bottom plots of Figure \ref{fig:jumps}. 

The ``sagging'' artifact correlates with an anomalously large dither pattern of amplitude roughly 10 arcmin, as opposed to the 1 arcmin spiral dither sequence. 
Dither pattern amplitudes for sagging cases are roughly distributed as a skew normal with mean $\sim6$ arcmin, 
standard deviation $\sim2$ arcmin and shape parameter $\sim3$ arcmin. 
Triangle-wave variations are superposed to a sinusoidal dither pattern throughout the duration of the visit and arise in spurts lasting 
the length of the spiral pattern cycle ($\sim120$ sec).
Unlike jumps or triangle-wave variability, sagging artifacts occur in both bands in the same visit, but not
necessarily with the same amplitude or sign.

To examine whether the sagging artifact is a detector-wide effect, we investigate other sources in the field during visits when a source from our sample 
exhibits sagging. 
We find that other sources are affected by the artifact during the observation,
however, sagging artifacts do not necessarily occur in the same way for all sources: some sources may increase and then decrease in flux, while
other sources may decrease then increase in flux over the same timespan. We attribute this reflection to local variations in detector response. 

To investigate possible correlations with observation parameters, 
we examine $\Delta \textrm{mag}$ against date of observation, detector radius and count rate for light curves showing
sagging artifacts in the second-to-bottom panel of Figure \ref{fig:jpc_labels}. 
We find little correlation between $\Delta \textrm{mag}$ and other quantities in Figure \ref{fig:jpc_labels} for either band. Sagging artifacts, as all other 
artifacts discussed thus far, occur in GALEX time-resolved photometry since a few months after the mission launched.
They happen rarely within the first 1500 days after launch, but occur more frequently afterwards.
Sagging cases take place more discretely in time than other artifacts
after the spike in incidence rate at 1500 days after launch in our limited sample. 

\begin{figure*}
\centering

\includegraphics[width=1.2in]{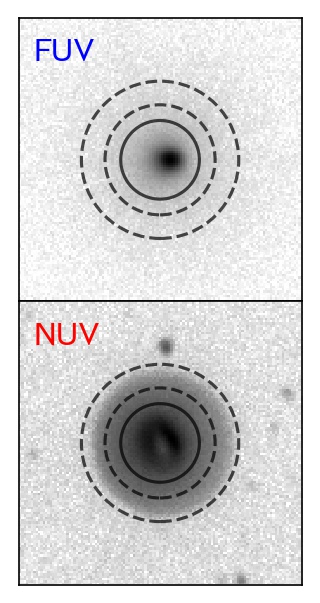}\includegraphics[width=5.25in]{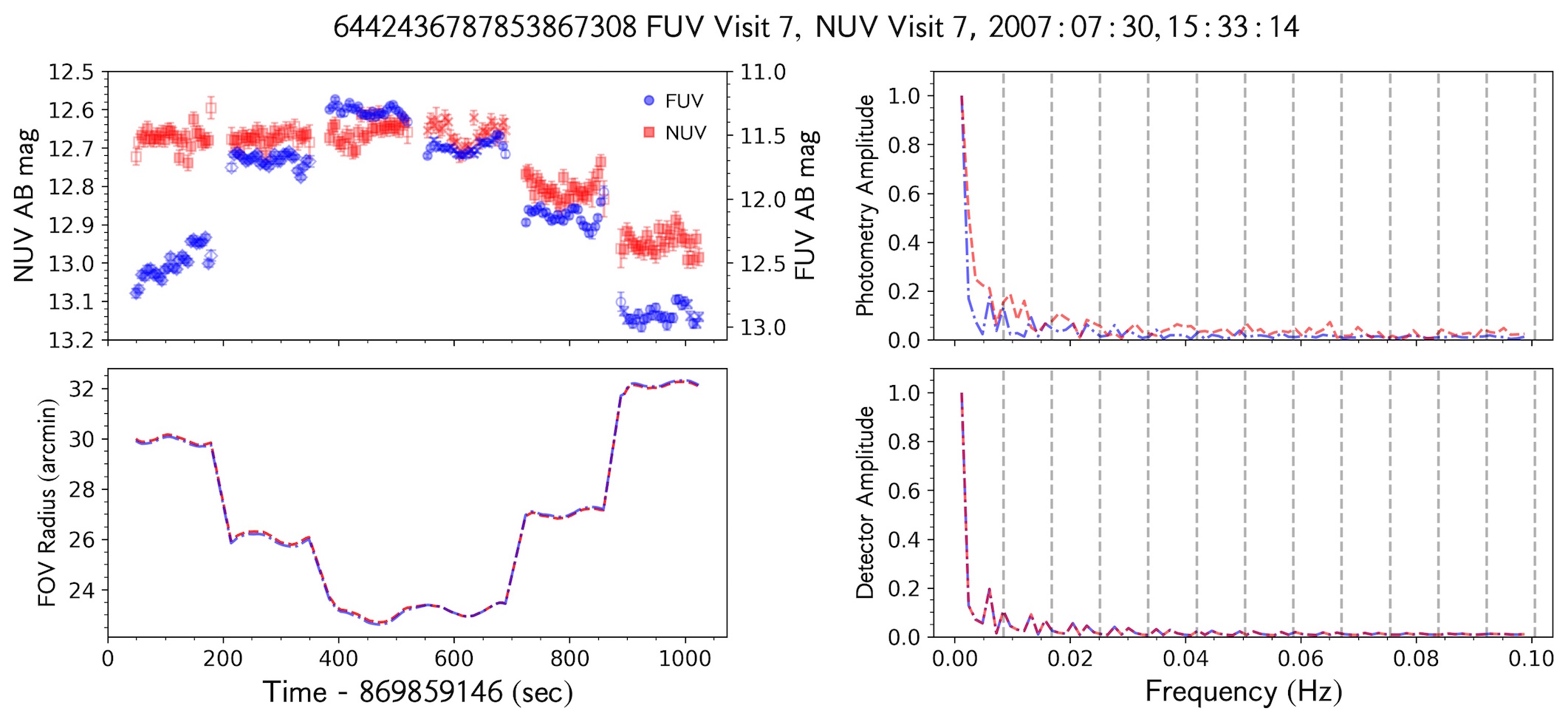}
\includegraphics[width=0.9in]{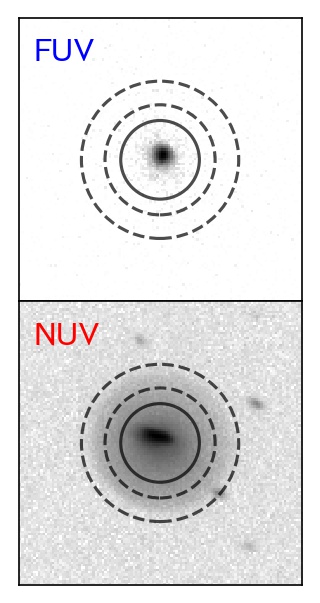} \includegraphics[width=3.25in]{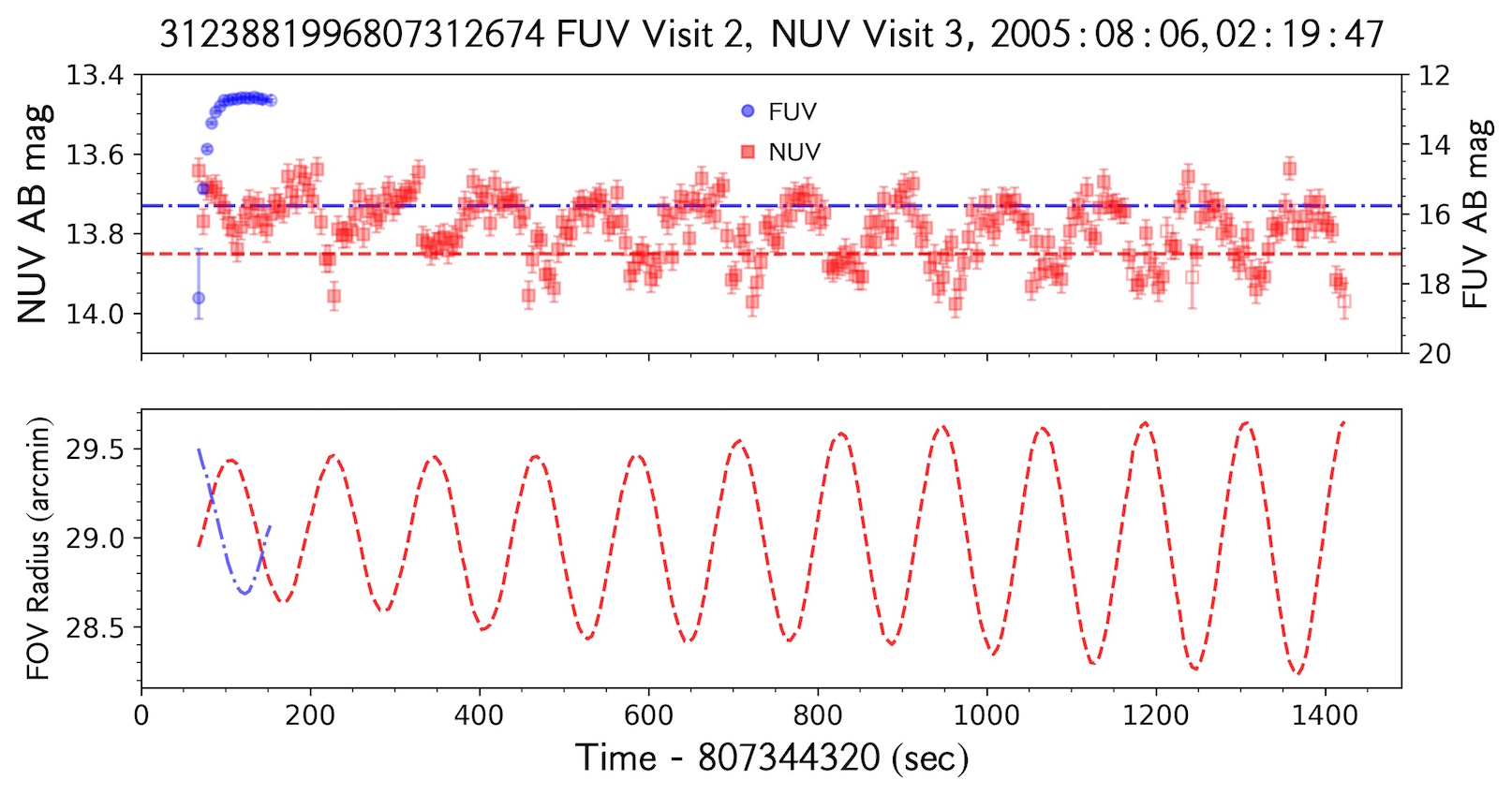} \includegraphics[width=2.75in]{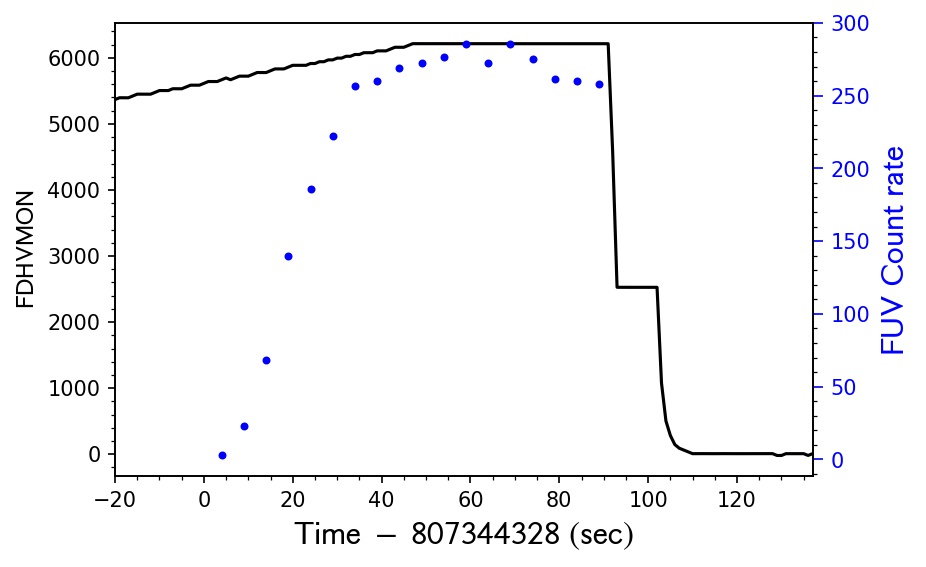} \\
\includegraphics[width=0.9in]{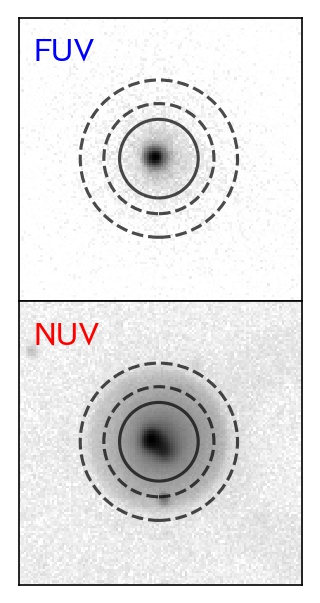} \includegraphics[width=3.25in]{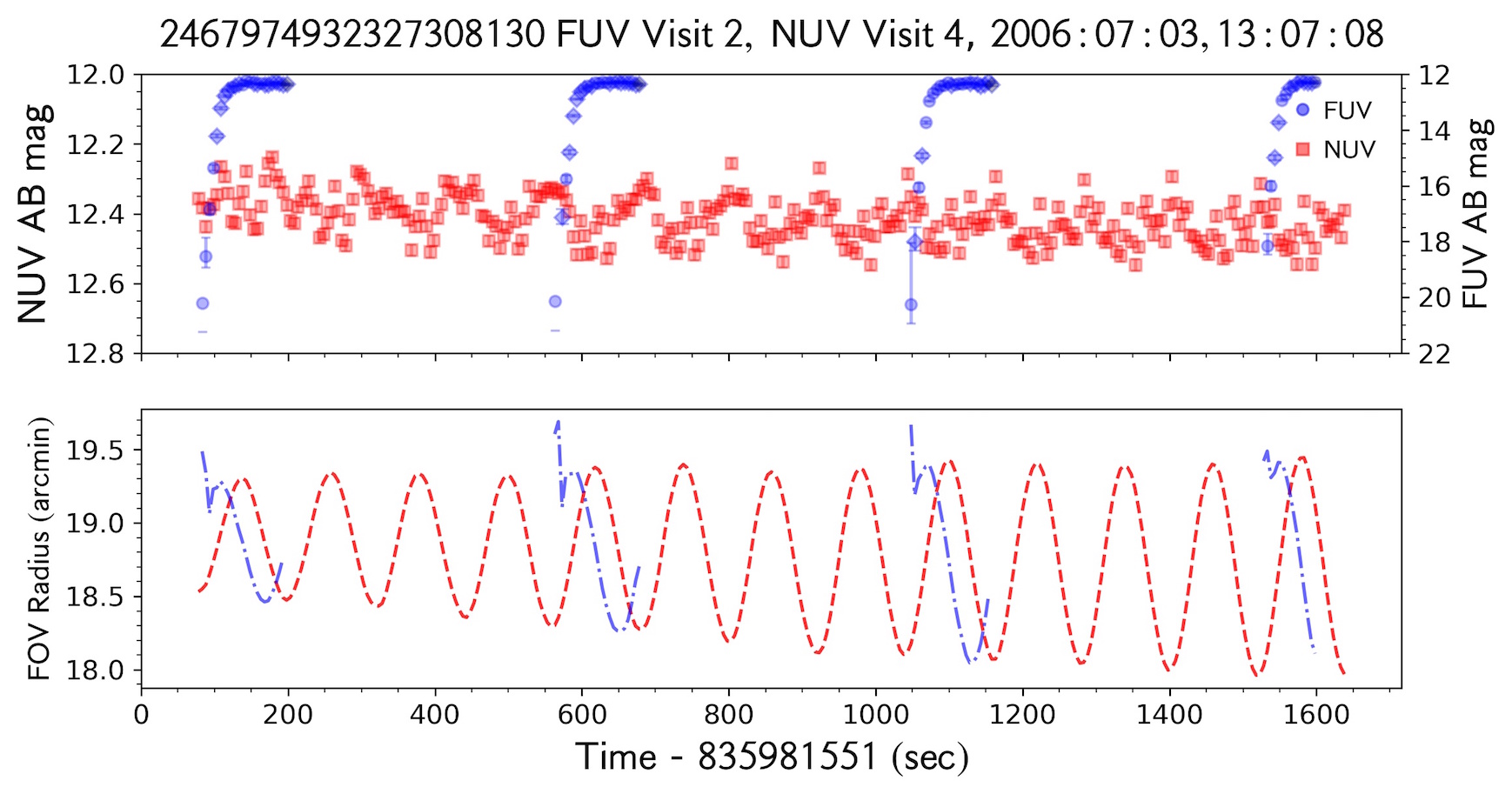} \includegraphics[width=2.75in]{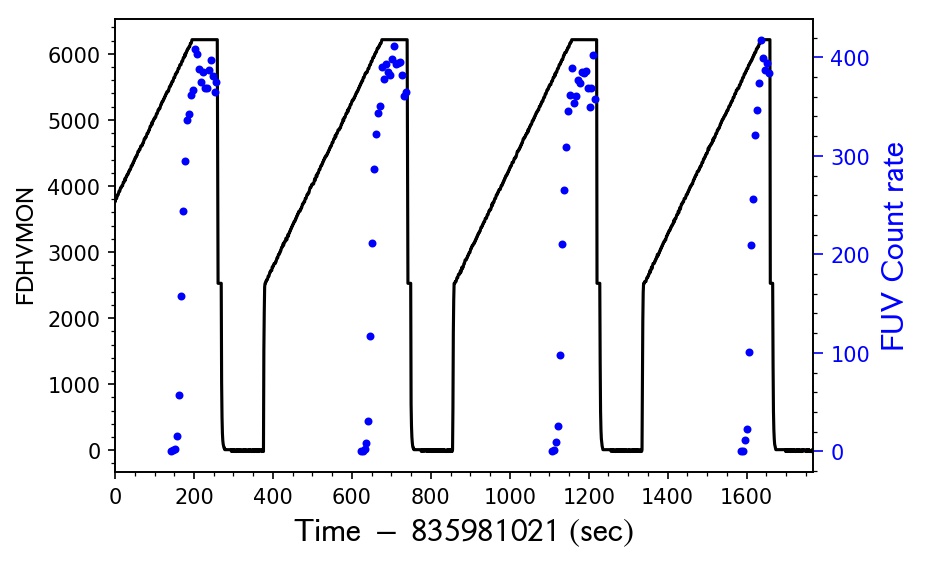}

\caption{Similar to Figures \ref{fig:triangles} and \ref{fig:jumps}, showing examples of the ``wedding cake'' version of the sagging artifact (top panels), 
which can produce $\Delta \textrm{mag} \geq 1.0$ mag in both bands or in one band (top plot, FUV).
The bottom rows show two examples of the rapid increase in FUV brightness.
Rapid rises in FUV brightness can arise at the start of visits and last less than 200 seconds (middle row) or recur a few times during the visit (bottom row). 
The right panels in the middle and bottom rows show the FUV detector voltage along with FUV count rate. FUV brightness 
correlates with the voltage cycling on and off.
We note that in all cases of this artifact, the FUV dither pattern is out of phase by roughly one third of a cycle with respect to the NUV dither pattern, 
suggesting a misalignment in the image reconstruction.
}
\label{fig:wedding_cake}
\end{figure*}

We also encountered 18 visits with a more extreme version of this artifact. In the top panel of Figure \ref{fig:wedding_cake} 
we show an example. This artifact is characterized by a dither pattern of amplitude $\gtrsim10$ arcmin, but instead of a spiral pattern
superposed to a sinusoidal dither, as with the sagging artifact, the dither sequence
resembles the tiers of a wedding cake. This ``wedding cake'' dither motion often leads to variations $\gtrsim 1.0$ mag in both bands during the visit. 
In the bottom panel of Figure \ref{fig:jpc_labels} we highlight these 18 visits with a black, dashed ellipse. 
All 18 incidences of this artifact occur on one day, 30 July 2007, roughly 1550 days after GALEX launch. 


Figure \ref{fig:jpc_labels}, second-to-bottom panel, shows that sagging artifacts occur at any {\tt fov\_radius}, and do not preferentially happen in one band. 
Figure \ref{fig:triangle_quad} shows {\it paths on the detector}, during the visit, for all sources in our sample, with 
those affected by the sagging artifact separated by band. Most dither patterns during sagging produce circular tracks on the detector.
Dither patterns with amplitude $\gtrsim15$ arcmin usually do not yield circular tracks, so the mean X, Y position may not be appropriate 
in calculating the artifact fraction in these visits. These cases are exceptionally rare though, and should not bias our artifact fraction measurements. 

Artifact fractions for sagging cases, plotted in the bottom row, right panel of Figure \ref{fig:quad_artifact},
are remarkably similar in both FUV and NUV.
Fractions fluctuate about a mean of $\sim15\%$ out to a {\tt fov\_radius} of 30 arcmin.

\subsection{Rapid Rises in FUV Brightness and Other Extreme Variations}
\label{subsec:rise}

We find 66 light curves with $\Delta \textrm{mag} \geq 1.0$ mag. Of these, 33 are due to a rapid ($\lesssim$ 30 sec) increase in FUV brightness, 
13 arise from a dither pattern similar to that which causes the 
sagging artifact, 18 are extreme cases of the sagging artifact, and 2 come from visits where the dither pattern oscillates wildly and does not cause sagging artifacts. 
We show in the bottom two rows in Figure \ref{fig:wedding_cake} two cases of the rapid rise in FUV brightness. 
In most cases, this artifact appears only at the beginning of the visit (bottom left panel) but it can recur a few times throughout 
the visit (bottom row). For all cases where this quick FUV increase appears, we note that the FUV dither pattern is out of phase with respect to the NUV 
dither pattern by about a third of a cycle.

{For the ``FUV rise'' examples shown in Figure \ref{fig:wedding_cake} we examined additional spacecraft parameters: detector temperature and voltage
(parameters {\tt FDTHVPS}, {\tt FDTLVPS}, {\tt FDHVMON}, {\tt HVNOM\_FUV} in spacecraft state files (extension -scst.fits)). 
FUV detector voltage readings correlate with FUV brightness and appear
to have been ``cycled" on and off during the observations. In Figure \ref{fig:wedding_cake}, bottom two rows, right panels, we plot
{\tt FDHVMON}, a measure of the FUV detector voltage \citep{morrissey06}, along with FUV count rate during the observation.
Count rates are high only when the voltage reaches the nominal level, and are not measured when the voltage is zero.
The parameters {\tt FDTHVPS} and {\tt FDTLVPS}, which correspond to 
detector temperature, have no influence on the count rate.

No correlation is observed with {\tt fov\_radius} or count rate (except for the fact that this 
artifact was found for sources in the nonlinear regime), but all instances of the FUV rise in our sample lie in a narrow range in observation date,
indicated by the blue dotted ellipse in the left panel, bottom row of Figure \ref{fig:jpc_labels}. 
Specifically, these artifacts occur $\sim850 - 1200$ days after the GALEX launch, i.e. August 2005 to July 2006.
This date range could allow us to constrain a potential cause for the FUV rise, although we must recall our small number statistics. 
\citet{morrissey07} report a FUV anomaly in 2005 in their Sec. 4.2, which resulted in the FUV being cycled on and off for short observation periods. 
We do not detect FUV rises in our sample after 2006.

\begin{figure*}
\centering
\includegraphics[width=5.in]{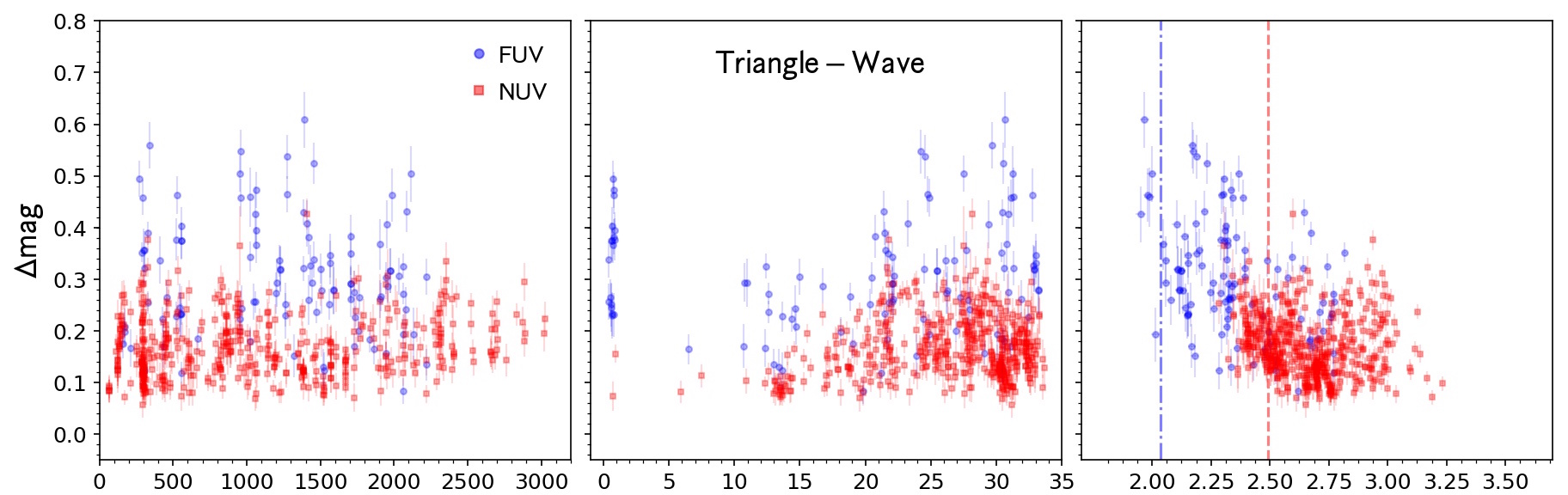}
\includegraphics[width=5.in]{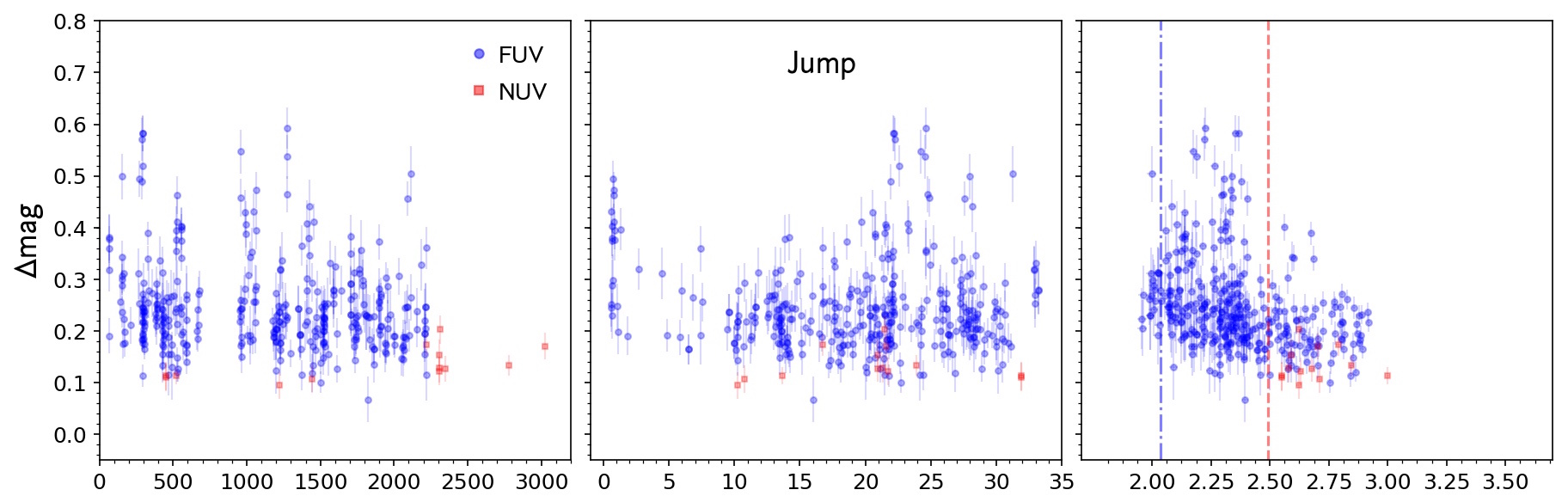}
\includegraphics[width=5.in]{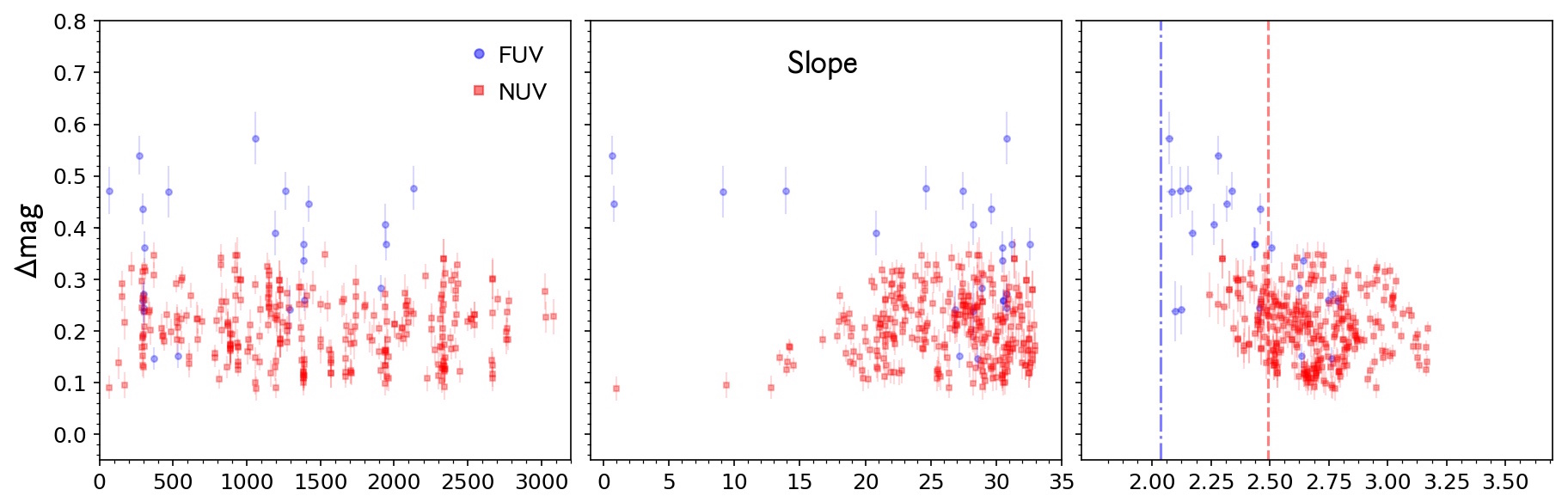}
\includegraphics[width=5.in]{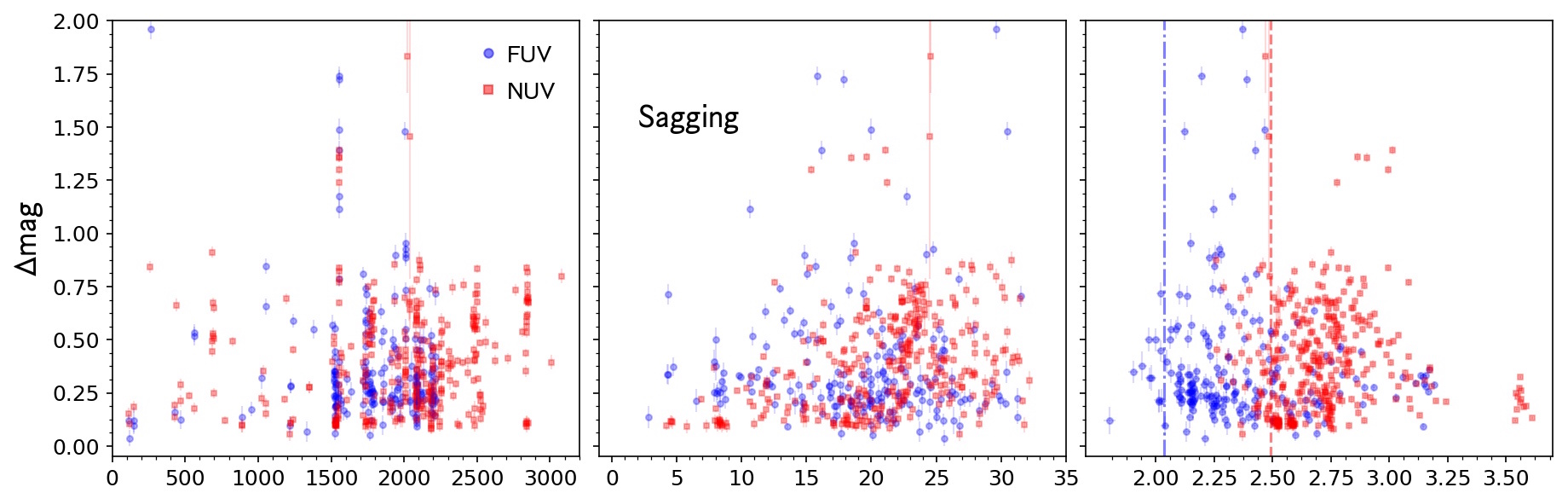}
\includegraphics[width=5.in]{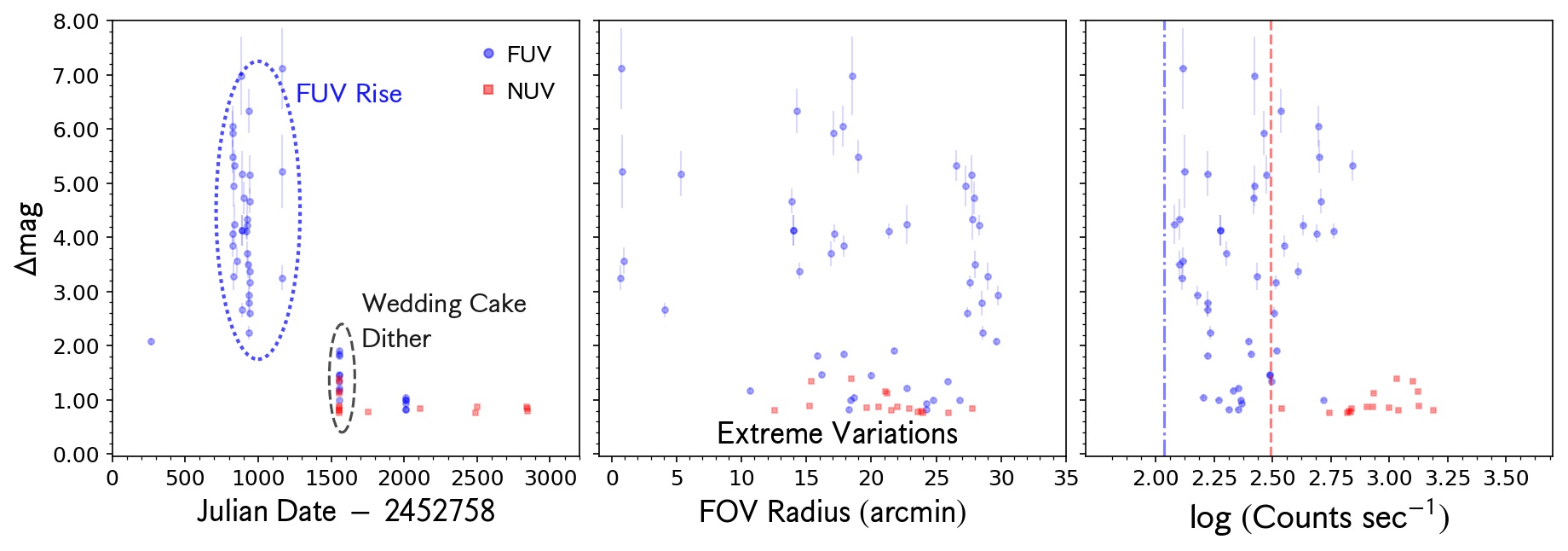}
\caption{Maximum variation in brightness, $\Delta \textrm{mag}$, as a function of (left to right) Julian Date since GALEX launch, 
mean detector radius, and average, background-subtracted count rate per visit, separated by band, for four types of artificial variations (top four rows)
and extreme variability (i.e. $\Delta \textrm{mag} \geq 1$, bottom row). 
Blue dash-dot and red dashed lines in the rightmost panels indicate the FUV and NUV nonlinear cutoffs, respectively.
Each of these four types of artifacts appeared within three months of the GALEX launch, and show little dependence on detector radius or count rate during the visit.
In the bottom left panel, all visits corresponding to rapid increases in FUV brightness are circled by the blue dotted ellipse, and 
all visits exhibiting the wedding cake dither pattern are circled by the black dashed ellipse. 
The wedding cake dither sequence is observed on only one date in our entire sample, 30 July 2007.
}
\label{fig:jpc_labels}
\end{figure*}

\begin{sidewaystable*}[h!]
    \centering
    \caption{Occurrence of artifacts as a function of {\tt fov\_radius} and quadrant.}
    \begin{tabular}{ccccccccccccccc}
    \hline
    Artifact & Quadrant & \multicolumn{2}{c}{${\tt fov\_radius}$\tablenotemark{a} $< 7$} & \multicolumn{2}{c}{$7 < {\tt fov\_radius} < 14$} & \multicolumn{2}{c}{$14 < {\tt fov\_radius} < 21$} & \multicolumn{2}{c}{$21 < {\tt fov\_radius} < 28$} & \multicolumn{2}{c}{$28 < {\tt fov\_radius} < 35$} & \multicolumn{2}{c}{Total} \\
     & & FUV & NUV & FUV & NUV & FUV & NUV & FUV & NUV & FUV & NUV & FUV & NUV \\
    \hline
   \multirow{4}{*}{Triangle-Wave} & 1 & 4 / 24 & 0 / 32 & 3 / 77 & 5 / 95 & 2 / 125 & 33 / 163 & 0 / 147 & 74 / 211 & 5 / 160 & 118 / 218 & 14 / 533 & 230 / 719\\
   & 2 & 0 / 25 & 1 / 30 & 3 / 53 & 9 / 66 & 7 / 139 & 20 / 159 & 13 / 167 & 106 / 252 & 20 / 217 & 125 / 285 & 43 / 601 & 261 / 792 \\
   & 3 & 11 / 61  & 1 / 84  & 4 / 50  & 5 / 64  & 2 / 102  & 15 / 132 & 25 / 122 & 51 / 215 & 19 / 159 & 31 / 231 & 61 / 494 & 103 / 726 \\
   & 4 & 7 / 37  & 1 / 52  & 3 / 51 & 3 / 68 & 3 / 117 & 23 / 165 & 3 / 134 & 58 / 227 & 0 / 173 & 102 / 229 & 16 / 512 & 187 / 741 \\
   Total & & 22 / 147 & 3 / 198 & 13 / 231 & 22 / 293 & 14 / 483 & 91 / 619 & 41 / 570 & 289 / 905 & 44 / 709 & 376 / 963 & 134 / 2140 & 781 / 2978 \\
    \hline
    \multirow{4}{*}{Jump} & 1 & 6 / 24 & 0 / 32 & 24 / 77 & 1 / 95 & 32 / 125 & 1 / 163 & 37 / 147 & 0 / 211 & 11 / 160 & 0 / 218 & 110 / 533 & 2 / 719\\
   & 2 & 1 / 25 & 0 / 30 & 19 / 53 & 0 / 66 & 21 / 139 & 0 / 159 & 49 / 167 & 0 / 252 & 28 / 217 & 0 / 285 & 118 / 601 & 0 / 792 \\
   & 3 & 12 / 61  & 0 / 84  & 7 / 50  & 1 / 64  & 12 / 102  & 0 / 132 & 24 / 122 & 1 / 215 & 14 / 159 & 0 / 231 & 69 / 494 & 2 / 726 \\
   & 4 & 9 / 37  & 0 / 52  & 13 / 51 & 1 / 68 & 28 / 117 & 2 / 165 & 15 / 134 & 4 / 227 & 0 / 173 & 2 / 229 & 65 / 512 & 9 / 741 \\
   Total & & 28 / 147 & 0 / 198 & 63 / 231 & 3 / 293 & 93 / 483 & 3 / 619 & 125 / 570 & 5 / 905 & 53 / 709 & 2 / 963 & 362 / 2140 & 13 / 2978 \\
    \hline
    \multirow{4}{*}{Slope} & 1 & 0 / 24 & 0 / 32 & 0 / 77 & 0 / 95 & 1 / 125 & 12 / 163 & 1 / 147 & 25 / 211 & 2 / 160 & 41 / 218 & 4 / 533 & 78 / 719\\
   & 2 & 0 / 25 & 0 / 30 & 0 / 53 & 0 / 66 & 1 / 139 & 7 / 159 & 1 / 167 & 41 / 252 & 11 / 217 & 53 / 285 & 13 / 601 & 101 / 792 \\
   & 3 & 3 / 61  & 1 / 84  & 2 / 50  & 1 / 64  & 1 / 102  & 6 / 132 & 4 / 122 & 46 / 215 & 3 / 159 & 19 / 231 & 13 / 494 & 73 / 726 \\
   & 4 & 1 / 37  & 0 / 52  & 0 / 51 & 3 / 68 & 0 / 117 & 12 / 165 & 0 / 134 & 30 / 227 & 0 / 173 & 40 / 229 & 1 / 512 & 85 / 741 \\
   Total & & 4 / 147 & 1 / 198 & 2 / 231 & 4 / 293 & 3 / 483 & 37 / 619 & 6 / 570 & 142 / 905 & 16 / 709 & 153 / 963 & 31 / 2140 & 337 / 2978 \\
    \hline
    \multirow{4}{*}{Sagging} & 4 & 4 / 24 & 6 / 32 & 14 / 77 & 18 / 95 & 15 / 125 & 27 / 163 & 13 / 147 & 35 / 211 & 2 / 160 & 6 / 218 & 48 / 533 & 92 / 719\\
   & 2 & 7 / 25 & 8 / 30 & 4 / 53 & 7 / 66 & 21 / 139 & 20 / 159 & 28 / 167 & 55 / 252 & 3 / 217 & 8 / 285 & 63 / 601 & 98 / 792 \\
   & 3 & 6 / 61  & 8 / 84 & 8 / 50  & 13 / 64  & 13 / 102  & 20 / 102 & 10 / 122 & 25 / 215 & 3 / 159 & 4 / 231 & 40 / 494 & 70 / 726 \\
   & 4 & 7 / 37 & 8 / 52  & 11 / 51 & 11 / 68 & 21 / 117 & 28 / 165 & 18 / 134 & 44 / 227 & 0 / 173 & 4 / 229 & 57 / 512 & 95 / 741 \\
   Total & & 24 / 147 & 30 / 198 & 37 / 231 & 49 / 293 & 70 / 483 & 95 / 619 & 69 / 570 & 159 / 905 & 8 / 709 & 22 / 963 & 208 / 2140 & 355 / 2978 \\
    \hline
    \end{tabular}
    \label{tab:artifact}
    \tablenotetext{a}{{\tt fov\_radius} in arcmin.}
\end{sidewaystable*}

\begin{figure*}
\centering
\includegraphics[width=3.25in]{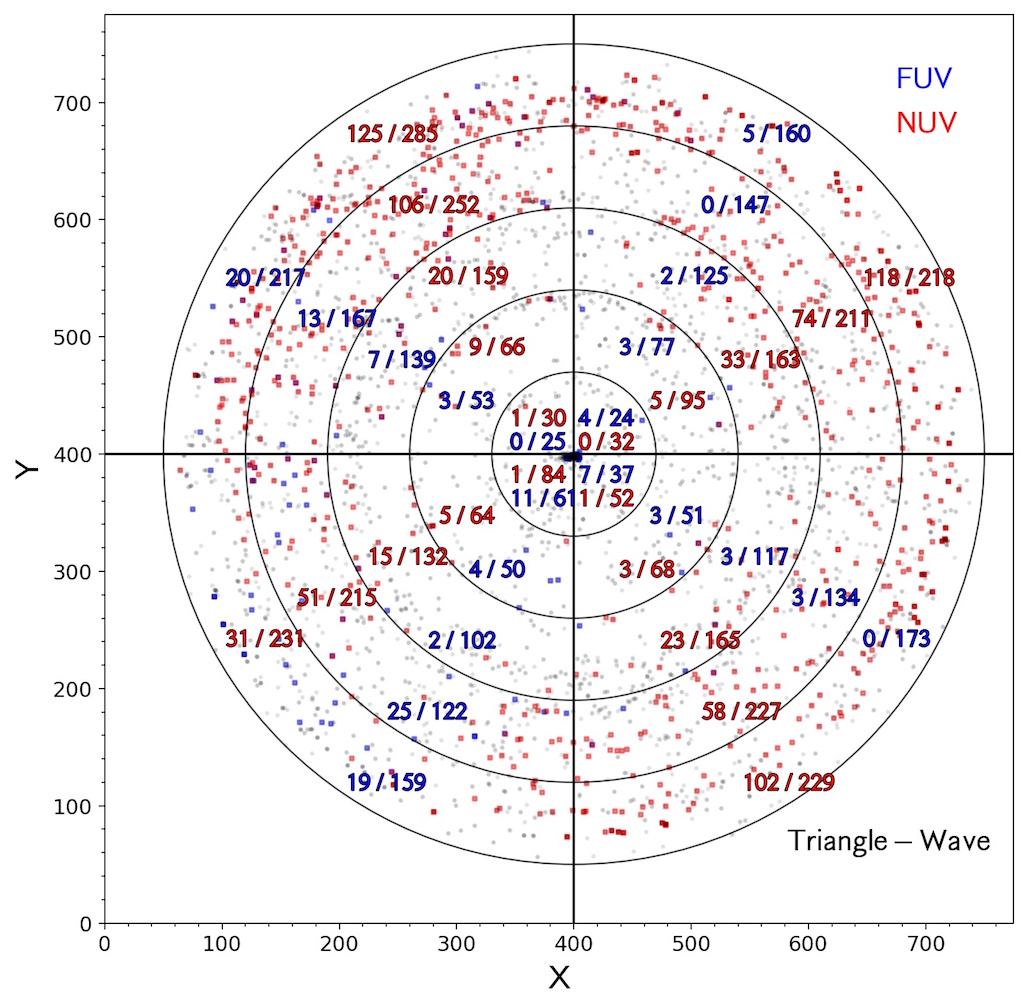}
\includegraphics[width=5.5in]{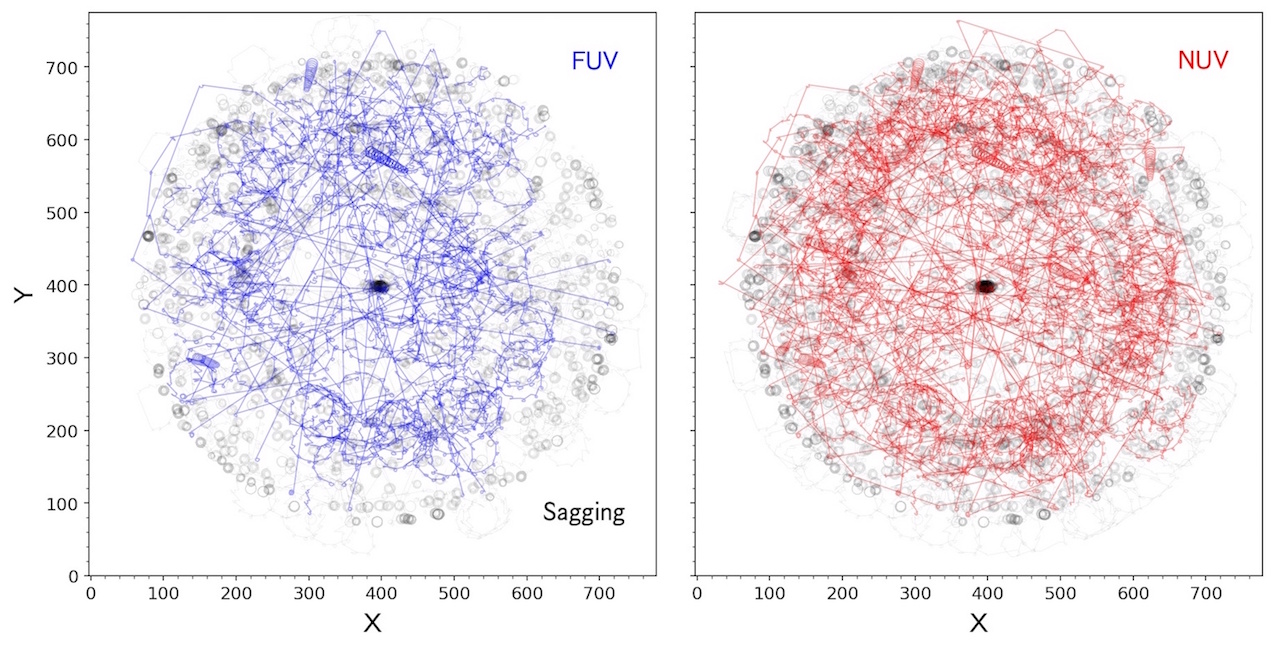}

\caption{
Positions of all sources on the detector (light gray points) and of those affected by triangle-wave artifacts (colored by band, top), 
Concentric circles 7 arcmin apart represent our five radial bins, which we further divide by quadrant. 
Numbers within each radial bin indicate the fraction of triangle-wave occurrence for that spatial bin. Coordinate values are in native \gphoton \ 
units. Triangle-wave artifacts more often occur at ${\tt fov\_radius} > 20$ arcmin in all quadrants except the third. 
{\it Paths on the detector} for all observations (light gray) and for sources in all observations affected by the sagging artifact, separated by band (bottom panels).
Typical dither patterns are 1 arcmin spirals throughout the visit and appear as small, 
gray annuli in this plot. Dither patterns during sagging light curves usually appear as serrated, wide ($\sim6$ arcmin)
circles, or lines streaking across the detector (``wedding cake'' artifacts). 
}
\label{fig:triangle_quad}
\end{figure*}

\begin{figure*}
\centering

\begin{tabular}{cc}
\includegraphics[width=3.in]{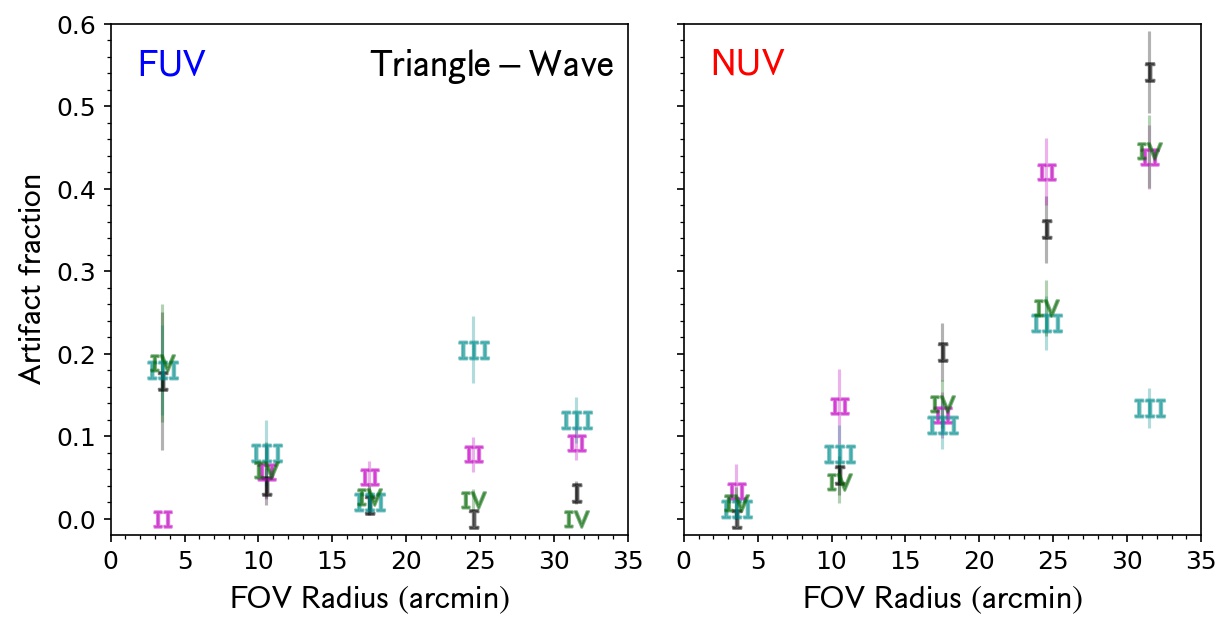}
&
\includegraphics[width=3.in]{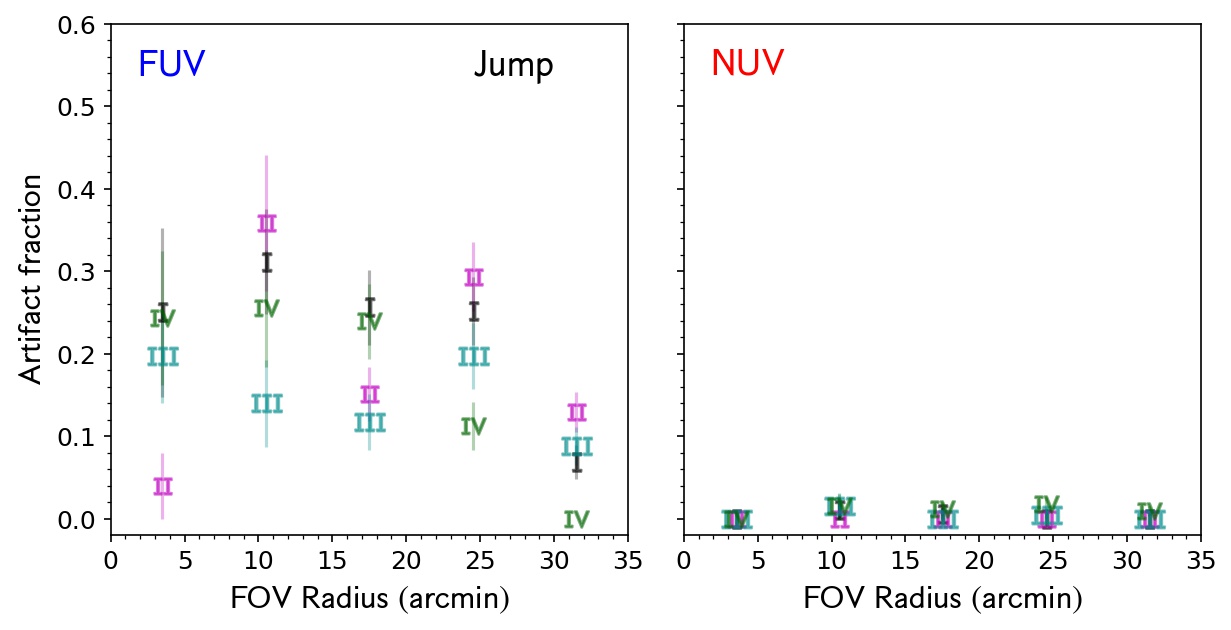} \\

\includegraphics[width=3.in]{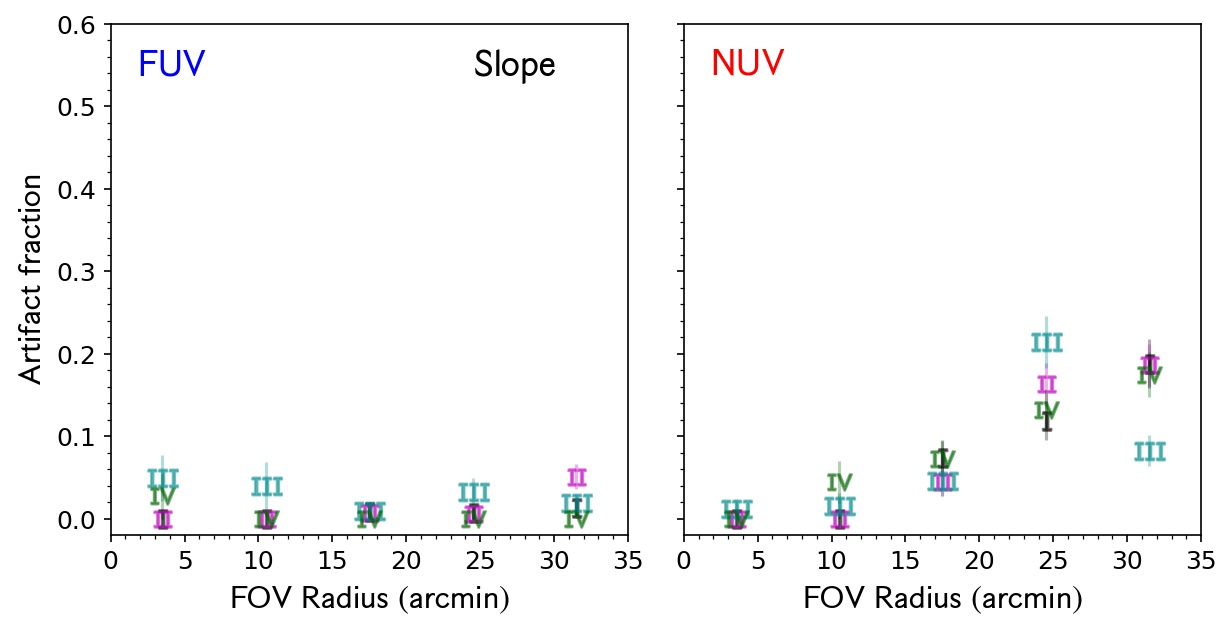}
&
\includegraphics[width=3.in]{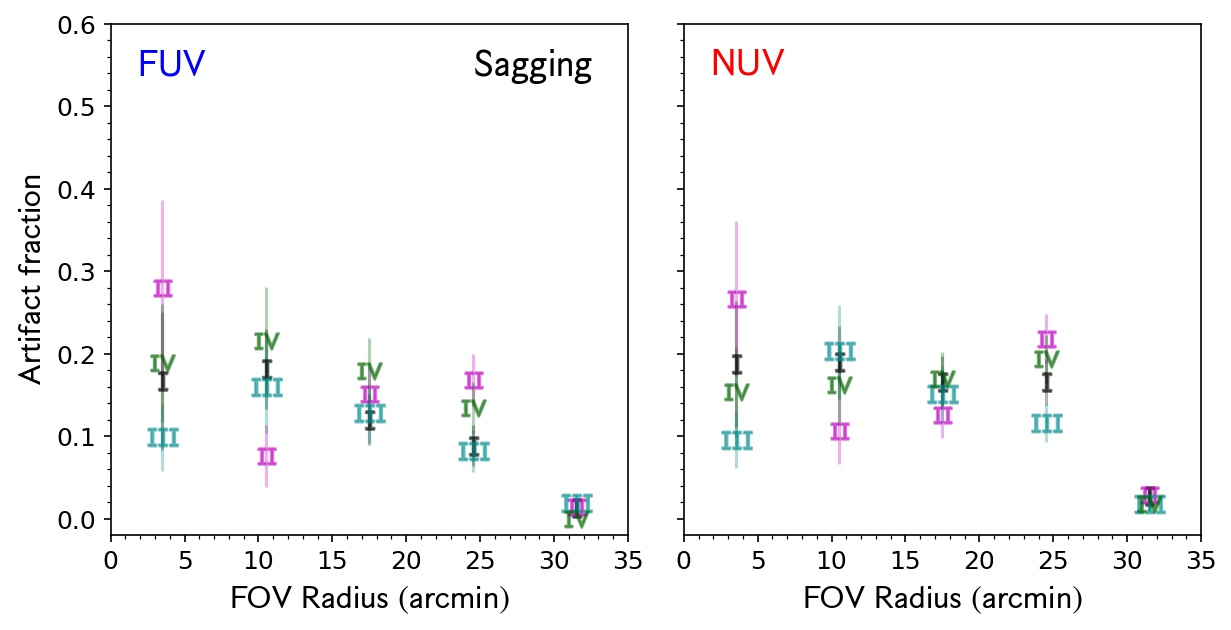}
\end{tabular}

\caption{
Fraction of light curves affected by artifacts, as a function of quadrant on the detector and {\tt fov\_radius}, separated by band. 
Quadrant fractions are indicated by roman numerals.
Fractions per artifact type, quadrant, and radial bin are listed in Table \ref{tab:artifact}.
Triangle waves and slopes are predominantly NUV artifacts and are more likely to occur at {\tt fov\_radius} $\geq 15$ arcmin, reaching fractions of
10 to 20 percent. Jumps are FUV artifacts and happen often at {\tt fov\_radius} $\leq 30$ arcmin, with 15 to 30\% of FUV visits affected.
Sagging artifacts occur roughly 15\% of the time in both bands at {\tt fov\_radius} $\leq 30$ arcmin.
}
\label{fig:quad_artifact}
\end{figure*}


\section{Discussion. Detection and Characterization of Artifacts}
\label{sec:detect_char}

In preparation for a comprehensive automatic search for variability across a large sample 
we have devised and tested methods to identify and eliminate spurious (instrumentally-induced) variability. 
We summarize the methodology below. 

The most common types of periodic variability (triangle-waves and sagging) in our sample are related to the dither motion during the visit.
The most prevalent non-periodic artifact, the jump, does not depend on the dither.
For the first case, one can discern 
artifacts by comparing the most significant periodicities between the 
time-resolved photometry and the detector periodicity.

We compute the Fourier Transform (FT) for each light curve and identify the peak 
frequencies in the FT of the light curve and the FT of the detector radius as a function of time. 

\subsection{Identifying Triangle-Wave and Jump Artifacts}

Triangle-wave variations in time-resolved photometry are strongly correlated with the dither pattern. When the triangle-wave is the dominant
variability throughout the light curve, the peak frequency of the light curve is the fundamental frequency of the dither sequence, or some harmonic thereof. 
In visits affected by the jump artifact, the largest change in magnitude
during the visit usually occurs at the time of the jump. As the jump only occurs once per observation, the largest power in the FT is found
in the lowest frequencies, corresponding to long-period variations. 
Slope artifacts, similar to jumps in behavior, have the highest power
at the lowest frequencies in their light curve FTs. The triangle-wave occurs so often alongside, however, that the FT of the photometry 
usually has more power at the dither pattern frequency than at frequencies corresponding to the inverse of the visit length. 
Sagging artifacts excite anomalously large variations in both the time-resolved photometry and detector radius of period roughly the visit length. 
The peak frequency for both the light curve and the detector radius is the inverse of the visit length or some integer multiple thereof for this artifact. 

\begin{figure*}
\centering
\includegraphics[width=4.in]{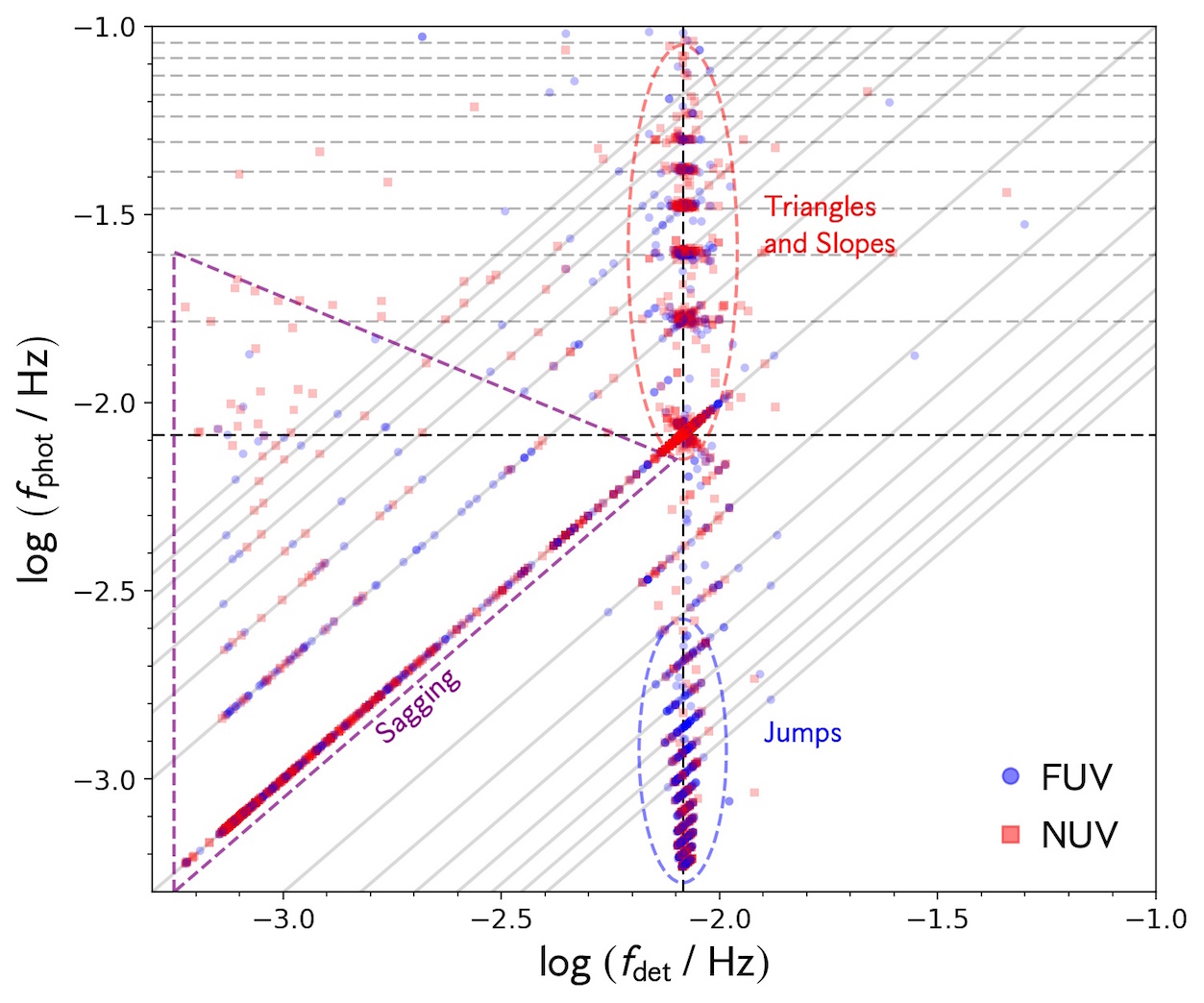}
\caption{Peak frequency derived from the dithering pattern, $f_{\textrm{det}}$, against peak frequency from photometry, $f_{\textrm{phot}}$, 
for light curves of length $\geq 200$ sec in our sample, colored by band. Dashed horizontal lines indicate integer multiples of the median $f_{\textrm{det}}$,
0.00823 Hz; the vertical dashed line indicates the median $f_{\textrm{det}}$ as well. Solid lines show where $f_{\textrm{phot}} = n f_{\textrm{det}}$, 
for $n$ an integer or integer reciprocal. Red and blue ellipses encircle visits identified as having triangle/slope and jump artifacts, respectively, and
the purple dashed triangle surrounds visits with sagging artifacts. The main types of artifacts in time-resolved GALEX photometry mostly separate into
three sections in $f_{\textrm{det}} - f_{\textrm{phot}}$ space.}
\label{fig:freq_compare}
\end{figure*}

Examples of FTs illustrating the paradigm above are shown in the top row of Figure \ref{fig:triangles} and top row of Figure \ref{fig:jumps} 
(for triangle-wave variations and jump artifacts), bottom row of Figure \ref{fig:triangles} (slopes), and bottom two rows of Figure \ref{fig:jumps} 
and top row of Figure \ref{fig:wedding_cake} (sagging). 

We now examine how well this method can identify artifacts for our entire sample. 
In Figure \ref{fig:freq_compare} we plot the peak frequency of the FT of the dither pattern, $f_{\textrm{det}}$, against the peak frequency 
of the light curve FT, $f_{\textrm{phot}}$, for all visits in our sample at least 200 sec long. With our visual classifications of these light curves
(see Section \ref{sec:artifacts}), we designate where triangle, slope, jump and sagging artifacts lie on this plot. Even without our 
designations, it is clear that our FT analysis neatly divides different periodicities using the coordinates $(f_{\textrm{det}}, f_{\textrm{phot}})$. 

We investigate the location of the artifact-affected cases on this diagram, in particular those related to the dither frequency and its harmonics. 
Dark dashed lines denote the median $f_{\textrm{det}}$ of the source in each visit. Roughly two thirds of triangles and slopes artifact 
$f_{\textrm{det}}, f_{\textrm{phot}}$ lie within 0.1 dex of the median $f_{\textrm{det}}$. The other third exhibit 
$f_{\textrm{phot}}$ that cluster tightly at harmonics of the median $f_{\textrm{det}}$, represented by light gray, dashed horizontal lines. 
About 80\% of the jump artifacts have peak frequencies that lie within the ellipse labeled ``Jumps'' in Figure \ref{fig:freq_compare} and
the remaining 20\% have $f_{\textrm{phot}}$ consistent with values observed for triangle/slope variabilities. This is likely the result 
of light curves showing both triangle and jump artifacts wherein the triangle periodicity overcomes the strength of the jump.
Sagging artifacts should exhibit $f_{\textrm{det}}, f_{\textrm{phot}}$ close to the inverse duration of the visit, as the anomalous
dither pattern in these cases completes one cycle over the length of the observation. Light gray, solid lines in Figure \ref{fig:freq_compare}
illustrate lines where $f_{\textrm{phot}} = n f_{\textrm{det}}$, where $n$ is either an integer or reciprocal of an integer. 
Nearly 90\% of all light curves showing ``sagging" lie on the 1:1 line. 
The minority of sagging cases not on the one-to-one line have $f_{\textrm{phot}} > f_{\textrm{det}}$ due to variations in spectral resolution.

Examples in Figures \ref{fig:triangles}, \ref{fig:jumps} and \ref{fig:wedding_cake} and the partitioning of points in 
Figure \ref{fig:freq_compare} strongly suggest that simply determining the peak frequencies of the light curve 
and dither pattern FTs efficiently identifies the different types of artificial variability we study in this paper. 
Though peak frequencies for each type of artifact exhibit some scatter within the regions we outline in Figure 
\ref{fig:freq_compare}, the methodology we describe here is useful for identifying artifacts in future large-scale studies using the \gphoton \ tool.

\subsection{Detecting and Characterizing Extreme Variations}
\label{subsec:ex_var}

We provide several ways to characterize light curves that show extreme variations not associated with flagged instrumental defects 
such as hotspots. 

The rise in FUV flux, described in Section \ref{subsec:rise}, is the largest artificial variation in our sample, with typical $\Delta \textrm{mag} \gtrsim 2$ mag 
(see bottom panel of Figure \ref{fig:jpc_labels}).
All 33 visits exhibiting this artifact in our sample occur within a nearly year-long time span, from August 2005 to July 2006. 
Future work concerning FUV observations in this time frame should take into account possible contamination by rapid increases 
in FUV brightness. We have observed in each case of the FUV rise a phase difference of a third of a cycle between the FUV and NUV 
dither patterns. Combining this phase difference with either the $\Delta \textrm{mag}$ or date of observation of the visit serves as a powerful indicator
of the FUV rise.

Extreme sagging artifact variations, which show the ``wedding cake'' dither pattern (discussed in Section \ref{subsec:rise}), have 
$1.0 < \Delta \textrm{mag} \lesssim 2.0$ mag and all 18 observations of this artifact occur on one day, 30 July 2007 (see bottom two panels 
of Figure \ref{fig:jpc_labels}). However, it is possible that our limited sample size prevented discovery of other cases of the wedding cake dither sequence. 
All affected cases have amplitude $\geq 10$ arcmin.
As nearly all cases of the sagging artifact have $\Delta \textrm{mag} < 1.0$ mag and dither pattern amplitude 4 - 8 arcmin, 
wedding cake visits can be identified by selecting potential cases of sagging, using the spectral analysis outlined in the 
previous section, and choosing visits with $\Delta \textrm{mag} > 1.0$ mag and dither motion amplitude $\geq 10$ arcmin. 

\subsection{Removing Non-periodic Artifacts}
\label{subsec:remove_artifact}

A range of artifacts can now be classified en masse using the techniques developed in the previous sections. These artifacts 
account for most of the variability in our sample, after removing hotspot and low response data points.
For periodic artifacts, namely triangle-wave variability, we rebin the light curve over an interval with the 
exact period of the dither pattern, then divide the count rate across the light curve by the interpolated mean count rate. 
Non-periodic artifacts, which do not depend on the dither pattern, are simple to model. 

\begin{figure}
\centering
\includegraphics[width=3.25in]{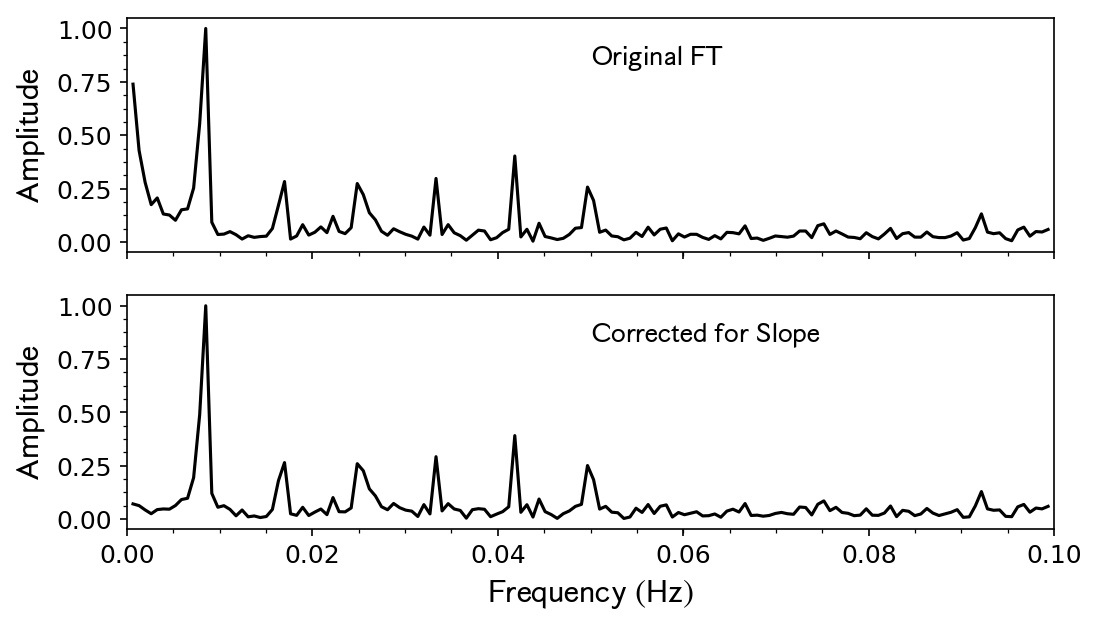}
\caption{Removing non-periodic artifacts. Top panel: original Fourier Transform of the light curve, showing a slope artifact 
(bottom panel of Figure \ref{fig:triangles}). Bottom panel: Fourier Transform of the same light curve after removing the slope
artifact using the method described in Section \ref{subsec:remove_artifact}. FTs are normalized to the peak amplitude.}
\label{fig:slope_corr}
\end{figure}

Results from our technique are shown in Figure \ref{fig:slope_corr}, using the slope artifact in the bottom panel of Figure
\ref{fig:triangles} as a test case. The original FT, shown in the top panel, has the highest power at the spiral dither 
frequency ($\approx 116^{-1}$ Hz), and a second-highest peak located at the inverse of the visit duration, due
to the slope artifact. After applying our procedure, the corrected FT, in the bottom panel of Figure \ref{fig:slope_corr}, 
shows no local maximum at the inverse visit duration, with the rest of the FT unaffected. Identical results are achieved
when testing on jump artifacts as well. 
This allows a cleaner identification and removal of triangle-wave artifacts from the FT. 

\section{Summary and Conclusions}
\label{sec:conclusion}

In this paper we have analyzed 5000 light curves of 304 bright ($m_{\rm{FUV}}, m_{\rm{NUV}} < 14$) and blue ($m_{\rm{FUV}} - m_{\rm{NUV}} < 0$) sources
using the database tool \gphoton. We inspected nearly 4000 light curves at least 200 seconds 
long with a time resolution of 5 seconds, and discovered several 
previously unreported artificial variations, after removing data points affected by hotspots, 
short integration times, low response and proximity to the detector edge, 
which are known causes of potential spurious variations and can be cleaned using the provided flags. 
Our results can be summarized as follows.

\begin{enumerate}

\item The most frequent artifacts we find are quasi-sinusoidal variations (``triangle-waves'') with periods $\sim120$ sec and amplitudes $\sim0.2$ mag.
These occur in either one or both bands but more often in NUV. They are caused by the spiral dither of the spacecraft pointing, which was used to minimize pixel-to-pixel
fluctuations. They can be easily identified by peaks in the light curve Fourier transform matching the dither frequency or its harmonics.
We attribute these to spatial inhomogeneities in detector response at the pixel scale.

\item Shifts in flux (``jumps'') and gradual changes in brightness (``slopes''), by up to a few tenths of a magnitude, 
and occur occasionally, the former more often in FUV and the latter more often in NUV. 

\item Sinusoidal-like variations, with periods equal to the duration of the observation, resembling a ``sagging'' or ``heaving'' of the source flux, 
with amplitudes $\sim 0.5$ mag, occur more rarely. These are accompanied by a dither motion with the same phase and orbit-long period. 
Other sources in the field exhibit the same artifact during the same observation,
but are not necessarily affected in the same way.
\end{enumerate}

We developed and tested a methodology to identify the artifacts and remove them from light curves using the
Fourier transforms of the light curve and the dither during an observation.
A future paper will be devoted to physical variations detected in GALEX time-resolved photometry (see examples in \citealt{bianchi18})
including this sample and to a fainter sample. 
The current sample is mostly in the non-linear regime although some of the artifacts are also seen at our faintest magnitude of 14.

\section*{Acknowledgements}
\noindent We are very grateful to the anonymous referee for the very careful reading and the many useful advices on this work. 
AD acknowledges fruitful discussions with Richard Anderson, Michael Busch, Kirill Tchernyshyov, and Anna Ulla.
AD and LB thank Patrick Morrissey for discussion about instrument configuration, Chase Million for help with \gphoton, 
and Gilles Fontaine and Paula Szkody for discussion. We acknowledge support from NASA Grant NNX17AF35G (16-ADAP16-0109).
This research made use of Astropy, a community-developed core Python package for Astronomy \citep{astro13}, as well as 
the SIMBAD database, operated at CDS, Strasbourg, France \citep{wenger00}.

{\it Software}: gPhoton \citep{million16}, Astropy \citep{astro13}, matplotlib \citep{hunter07}, and SciPy \citep{jones01}


\begin{thebibliography}{}

\bibitem[Astropy Collaboration(2013)]{astro13}
Astropy Collaboration, Robitaille, T., Tollerud, E. J., et al. 2013, A\&A, {\bf 558}, A33

\bibitem[Bianchi(2009)]{bianchi09}
Bianchi, L. 2009, Ap\&SS, {\bf 320}, 11

\bibitem[Bianchi et al.(2011a)]{bianchi11a}
Bianchi, L., Herald, J., Efremova, B., et al. 2011, Ap\&SS, {\bf 335}, 161

\bibitem[Bianchi(2014)]{bianchi14l}
Bianchi, L. 2014, Ap\&SS, {\bf 354}, 103

\bibitem[Bianchi et al.(2014)]{bianchi14}
Bianchi, L., Conti, A., \& Shiao, B. 2014, AdSpR, {\bf 53}, 973

\bibitem[Bianchi et al.(2017)]{bianchi17}
Bianchi, L., Shiao, B., \& Thilker, D. 2017, ApJS, {\bf 230}, 24

\bibitem[Bianchi et al.(2018)]{bianchi18}
Bianchi, L., de la Vega, A., Shiao, B., \& Bohlin, R. 2018, Ap\&SS, {\bf 363}, 56

\bibitem[Boudreaux et al.(2017)]{boudreaux17}
Boudreaux, T., Barlow, B. N., Fleming, S., W., et al. 2017, ApJ, {\bf 845}, 171 

\bibitem[Browne et al.(2009)]{browne09}
Browne, S. E., Welsh, B. Y., \& Wheatley, J. M. 2009, PASP, {\bf 121} 879

\bibitem[Conti et al.(2014)]{conti14}
Conti, A., Bianchi, L., Chopra, N., et al. 2014, AdSpR, {\bf 53}, 967

\bibitem[Costa et al.(2008)]{costa08}
Costa, J. E. S., Kepler, S. O., Winget, D. E., et al. 2008, A\&A, {\bf 477}, 627

\bibitem[Davenport et al.(2018)]{davenport18}
Davenport, J. R. A., Covey, K. R., Clarke, R. W., et al. 2018, ApJ, {\bf 853}, 130

\bibitem[Gezari et al.(2013)]{gezari13}
Gezari, S., Martin, D. C., Forster, K., et al. 2013, ApJ, {\bf 766}, 60

\bibitem[Hunter(2007)]{hunter07}
Hunter, J. D. 2007, Computing in Science and Engineering, {\bf 9}, 90

\bibitem[Jones et al.(2001)]{jones01}
Jones, E., Oliphant, T., Peterson, P., et al. 2001, SciPy: Open source scientific tools for python


\bibitem[Martin et al.(2005)]{martin05}
Martin, D. C., Fanson, J., Schiminovich, D., et al. 2005, ApJL, {\bf 619}, L1

\bibitem[Million et al.(2016)]{million16}
Million, C., Fleming, S. W., Shiao, B. et al. 2016, ApJ, {\bf 833}, 292 (M16)

\bibitem[Morrissey et al.(2005)]{morrissey05}
Morrissey, P., Schiminovich, D., Barlow, T. A. et al. 2005, ApJL, {\bf 619}, L7

\bibitem[Morrissey (2006)]{morrissey06}
Morrissey, P. 2006, ?GALEX Detector Flight Operations Guide,? (GAL-CIT-329v7b).

\bibitem[Morrissey et al.(2007)]{morrissey07}
Morrissey, P., Conrow, T., Barlow, T. A. et al. 2007, ApJS, {\bf 173}, 2, 682 (M07)


\bibitem[Robinson et al.(2005)]{robinson05}
Robinson, R. D., Wheatley, J. M., Welsh, B. Y., et al. 2005, ApJ, {\bf 633}, 447

\bibitem[Tucker et al.(2018)]{tucker18}
Tucker, M. A., Fleming, S. W., Pelisoli, I., et al. 2018, MNRAS, {\bf 475}, 4768

\bibitem[Welsh et al.(2006)]{welsh06}
Welsh, B. Y.,  Wheatley, J. M., Browne, S. E., et al. 2006, A\&A, {\bf 458}, 921

\bibitem[Welsh et al.(2007)]{welsh07}
Welsh, B. Y.,  Wheatley, J. M., Seibert, M. et al. 2007, ApJS, {\bf 173}, 673

\bibitem[Welsh et al.(2011)]{welsh11}
Welsh, B. Y.,  Wheatley, J. M., \& Neil, J .D. 2011, A\&A, {\bf 527}, 15

\bibitem[Wenger et al.(2000)]{wenger00}
Wenger, M., Ochsenbein, F., Egret, D. et al. 2000, A\&AS, {\bf 143}, 9

\bibitem[Wheatley et al.(2008)]{wheatley08}
Wheatley, J. M., Welsh, B .Y., \& Browne, S. E. 2008, AJ, {\bf 136}, 259

\bibitem[Wheatley et al.(2012)]{wheatley12}
Wheatley, J. M., Welsh, B .Y., \& Browne, S. E. 2012, PASP, {\bf 124}, 552


\end{thebibliography}
\end{document}